\RequirePackage{silence}
\documentclass[aps, pre, onecolumn, nofootinbib, notitlepage, groupedaddress, amsfonts, amssymb, amsmath, longbibliography, 11pt]{revtex4-2}
\usepackage{CJK}
\usepackage{graphicx}
\usepackage{hyperref}
\usepackage{xcolor}
\hypersetup{
    colorlinks,
    linkcolor={red!50!black},
    citecolor={blue!50!black},
    urlcolor={blue!80!black}
}
\usepackage{bm}
\usepackage{natbib}
\usepackage{longtable}
\usepackage{enumitem}
\setlist[description]{leftmargin=\parindent,labelindent=\parindent}

\usepackage{siunitx}

\usepackage{longtable}
\usepackage[caption=false]{subfig}

\begin{document}

\begin{CJK*}{GB}{}

\author{Guangrui Sun}
\email{mpesung@nus.edu.sg}
\affiliation{Department of Mechanical Engineering, National University of Singapore, 9 Engineering Drive 1, 117575 Singapore}
\author{J. Andrzej Domaradzki}
\affiliation{Department of Aerospace \& Mechanical Engineering, University of Southern California, Los Angeles, California 90089 USA}

\title{\Large Large eddy simulations of turbulent channel flow based on interscale energy transfer}

\begin{abstract}
\centering\begin{minipage}{\dimexpr\paperwidth-8.5cm}
\vspace*{1cm}

\hspace*{1em}
A previously developed modeling procedure for large eddy simulations (LESs) is extended to allow physical space implementations for inhomogeneous flows. The method is inspired by the well-established theoretical analyses and numerical investigations of homogeneous, isotropic turbulence. A general procedure that focuses on recovering the full subgrid scale (SGS) dissipation from resolved fields is formulated, combining the advantages of both the structural and the functional strategy of modeling. The interscale energy transfer is obtained from the test-filtered velocity field, corresponding subfilter scale (SFS) stress or, equivalently, the similarity model is used to compute the total SGS dissipation. The energy transfer is then cast in the form of eddy viscosity, allowing it to retain the desired total SGS dissipation and making the method numerically robust as an automatic step of backscatter control. The method is capable of providing a proper amount of total energy dissipation in actual, low resolution LES runs. The new approach is general and self-contained, working well for different filtering kernels, Reynolds numbers, and grid resolutions.

\end{minipage}
\end{abstract}
\maketitle
\end{CJK*}



\section{Introduction}

The governing equations in the present study are the Navier-Stokes equations for incompressible flow ($\nabla \cdot \bar{\bm{u}} = \frac{\partial \bar{u}_i}{\partial x_i} = 0$) defined in Cartesian coordinates
\begin{equation}
  \frac{\partial \bar{u}_i}{\partial t} + \frac{\partial}{\partial x_j}(\bar{u}_i \bar{u}_j + \tau_{ij}^{\textsc{sgs}}) = - \frac{1}{\rho} \frac{\partial \bar{p}}{\partial x_i} + \frac{\partial 2 \nu \bar{S}_{ij}}{\partial x_j},
  \label{eq:ns}
\end{equation}
where $\bm{u} = (u_1,u_2,u_3)$ is the velocity field, $x_1,x_2,x_3$ represent streamwise ($x$), vertical ($y$) and spanwise ($z$) coordinates, respectively, $p$ is the pressure, $\rho$ is the density, and $\nu$ is the kinematic viscosity. Overbar $\bar{\enspace}$ denotes implicit filtering due to LES grid. In (\ref{eq:ns})
\begin{equation}
  \tau_{ij}^{\textsc{sgs}} = \overline{u_i u_j} - \bar{u}_i \bar{u}_j, \quad \bar{S}_{ij} = \frac{1}{2}\left(\frac{\partial \bar{u}_i}{\partial x_j} + \frac{\partial \bar{u}_j}{\partial x_i}\right),
\end{equation}
are the SGS stress and the resolved strain rate, respectively. The former is an additional unknown that needs to be modeled.

In this work, we focus on maintaining an appropriate amount of total energy transfer by proposing a general SGS modeling procedure. The kinetic energy equation is obtained by multiplying (\ref{eq:ns}) by $\bar{u}_i$
\begin{equation}
  \frac{\partial \bar{E}}{\partial t} + \frac{\partial \left(\bar{E} \bar{u}_i \right)}{\partial x_i} + \frac{\partial \bar{u}_i \tau_{ij}^{\textsc{sgs}}}{\partial x_j} + \frac{1}{\rho} \frac{\partial \left( \bar{p} \bar{u}_i \right)}{\partial x_i} - \frac{\partial}{\partial x_i} \left(\nu \frac{\partial \bar{E}}{\partial x_i} \right) + \nu \frac{\partial \bar{u}_j}{\partial x_i} \frac{\partial \bar{u}_j}{\partial x_i} - \tau_{ij}^{\textsc{sgs}} \bar{S}_{ij} = 0,
  \label{eq:kesgs}
\end{equation}
where the resolved kinetic energy $\bar{E} = (1/2) \bar{u}_i \bar{u}_i$. The last two terms in the equation define the viscous dissipation $\varepsilon_\nu$ and the SGS dissipation $\varepsilon_{\textsc{sgs}}$, respectively.

According to the review of \citet{Domaradzki_2002b}, traditional SGS modeling approaches can be classified into three main categories: the eddy viscosity models, the similarity models and the mixed models which combine the former two expressions. Perhaps the most widely used group among them is the eddy viscosity models, which is based on the Boussinesq hypothesis that models turbulent energy transfer by analogy with the molecular diffusion
\begin{equation}
  \tau_{ij}^{\textsc{sgs}} - \frac{1}{3}\tau_{kk}^{\textsc{sgs}} \delta_{ij} = -2 \nu_t (\bar{S}_{ij} - \frac{1}{3}\bar{S}_{kk} \delta_{ij}),
  \label{eq:nutp}
\end{equation}
where $\nu_t$ is the eddy viscosity, which is allowed to vary in space and time. For incompressible flow, the trace $\bar{S}_{kk}$ is combined with the pressure term and only the deviatoric part is modeled.

Various SGS modeling procedures are typically evaluated by performing LES for different physical conditions and adjusting model constants iteratively until the best agreement with appropriate benchmarks is reached. In practice it is found that a crucial property of good SGS models is their ability to provide adequate SGS dissipation; conversely, models predicting inadequate SGS dissipation fail in actual simulations, irrespective of even a good performance in {\it a priori} tests. This observation suggests shifting the focus of SGS modeling from a functional form of a model (such as the Smagorinsky or similarity expressions) to determining the SGS transfer for a given LES field. Such a change of modeling focus was originated by \citet{Kraichnan_1976} who employed an analytical theory of turbulence, Test Field Model (TFM), to compute the energy transfer across a wavenumber cutoff $k_c$ between large scales ($k \leq k_c$) and subgrid scales ($k>k_c$), i.e., the SGS energy transfer, for isotropic turbulence in the inertial range. The theoretically computed SGS transfer can be cast in a form of the spectral eddy viscosity and used directly as a SGS model in LES of isotropic turbulence performed using Fourier spectral methods (see \citet{Lesieur_1996}). More recently the approach employing a computed SGS energy transfer as a modeling tool for actual LESs was extended by \citet{Domaradzki_2021a,Domaradzki_2021b,Domaradzki_2022a,Domaradzki_2022b} for the simulations of homogeneous isotropic turbulence (HIT). The main development is that the SGS transfer is obtained not from the theory but directly from evolving LES fields. The methodology is informed by an extensive literature on the subject of interscale energy transfer in turbulence obtained through analyses of direct numerical simulations (DNS) as outlined by \citet{Domaradzki_1990,Domaradzki_1994,Domaradzki_1995}, among others.
Moreover, in \citep{Domaradzki_2021a,Domaradzki_2021b,Domaradzki_2022a,Domaradzki_2022b} it was shown how the detailed SGS energy transfer among resolved scales obtained directly from the evolving LESs velocity fields can be used as a self-contained SGS model. Specifically, the SGS energy transfer among resolved scales and its wavenumber distribution is computed from LES fields at each time step and, following \citet{Kraichnan_1976}, cast in the form of a spectral eddy viscosity. Such a computed eddy viscosity is then modified to make it consistent with two known asymptotic properties of the energy flux in the inertial range and used in the eddy viscosity term added to the Navier-Stokes spectral solver as a SGS modeling term. Note that in this approach SGS modeling is accomplished without need for explicit expressions of the analytical theories or any other classical SGS models. While the spectral eddy viscosity is introduced in such an approach to SGS modeling, the primary physical quantity is the energy transfer across a wavenumber cutoff $k_c$ and the eddy viscosity is merely a derived quantity. Effectively, the procedure allows self-contained LESs without use of extraneous SGS models, or equivalently, at each time step the model is obtained from a simulated field itself and asymptotic properties of the energy flux in the inertial range.

The previous modeling procedure is defined entirely in spectral space. In this work we propose and investigate an application of this promising SGS modeling approach based on interscale energy transfer to LES of turbulent channel flow. The presence of inhomogeneous direction prevents direct application of spectral analysis employed for isotropic turbulence and the physical interpretation of energy transfer among scales is less clear. Nevertheless, as outlined in \citet{Domaradzki_2021a}, the same concepts can still be used in the context of test-filtered velocity fields in the physical space. We show that the subfilter scale dissipation computed from a resolved velocity field, can be used to obtain the total SGS dissipation for the full field. The subfilter scale stress (see e.g. \citep{Berselli_2006,Bull_2016}) is equivalent to the well-known similarity model \citep{Bardina_1980,Bardina_1983}, so the favorable properties of a structural-type model \citep{Sagaut_2006} can also be retained. Following the previous approach for HIT, the energy dissipation computed from resolved scales is recast to an eddy viscosity form, which is then used to estimate the SGS dissipation for the full LES fields. The new method is numerically accurate and robust. We show that the recasting procedure itself can be considered as a means of backscatter control that overcomes the numerical issues from the original similarity models. The inverse energy cascade is still allowed but greatly decreased, so that none of the present runs suffer from numerical instabilities. In addition, as it will be discussed in the following sections, the specific implementations are inspired by the theoretical analyses and well-established results for HIT \citep{Domaradzki_2021a,Domaradzki_2021b,Domaradzki_2022a,Domaradzki_2022b}, making it safe to extend the approach to wall-bounded flows. For various Reynolds numbers and filtering functions, we show that the new method can generate satisfactory results, without a deliberate tuning of the modeling coefficient. The model again confirms that maintaining the correct total SGS energy dissipation is a necessary condition for accurate large eddy simulations.

The rest of the paper is organized as follows. In Sec. \ref{sec:me}, we summarize the theoretical foundations of LES modeling based on the concept of interscale energy transfer. The extension to physical space implementation and corresponding numerical methods are presented in Sec. \ref{sec:nm}. A thorough analysis for turbulent channel flow for frictional Reynolds numbers $Re_\tau=180$, 1000, and 2000 are discussed in Sec. \ref{sec:res}. Finally, the main conclusions are drawn in Sec. \ref{sec:conc}.

\section{Methodology of the spectral eddy viscosity}
\label{sec:me}

In this section, for completeness, we briefly review the spectral space implementation of our previous modeling procedure based on the interscale energy transfer \citep{Domaradzki_2021a,Domaradzki_2021b}. The main features and observations from the large eddy simulations of homogeneous, isotropic turbulence offer a solid theoretical background, which will provide direct guidance in the extension of the approach to inhomogeneous flows.

In order to evaluate energy transfer among different scales in isotropic turbulence, it is natural to perform analyses in spectral space. The spectral LES energy equation for scales $k\leq k_c$ is obtained by defining first energy transfer $T^<(k|k_c)$ among resolved modes, where the notation signifies that only modes satisfying the inequality $k \leq k_c$, i.e. scales that are fully known in LES with the cutoff wavenumber $k_c$, are retained in computing $T^<(k|k_c)$. The complete spectral energy equation can then be rewritten for LES scales $k\leq k_c$ as follows
\begin{equation}\label{eq:SGS_en}
  \frac{\partial}{\partial t} E^<(k|k_c)=T^<(k|k_c)+T_{\textsc{sgs}}(k|k_c)-2 \nu k^2
  E^<(k|k_c), \; k \leq k_c,
\end{equation}
where the SGS energy transfer term is
\begin{equation}\label{eq:sgs}
  T_{\textsc{sgs}}(k|k_c)=T(k)-T^<(k|k_c), \; k \leq k_c,
\end{equation}
where $T(k)$ is the full nonlinear energy transfer computed using all modes up to the smallest Kolmogorov scale, including resolved and subgrid scales in LES.

Following \citet{Kraichnan_1976}, the SGS spectral energy equation can be formally rewritten as
\begin{equation}\label{eq:SGS_en1}
  \frac{\partial}{\partial t} E^<(k|k_c)=T^<(k|k_c)-
  2 \nu_t(k|k_c) k^2 E^<(k|k_c)-2 \nu k^2 E^<(k|k_c)
\end{equation}
where the SGS energy transfer is expressed in the same functional form as the molecular dissipation term by introducing the theoretical, effective eddy viscosity
\begin{equation}
\nu_t(k|k_c) = -\frac{T_{\textsc{sgs}}(k|k_c)}{2k^2E^<(k|k_c)}.
\label{eq:nu_th}
\end{equation}
As stressed in the Introduction, the eddy viscosity is a derived quantity, obtained from the primary physical quantity which is the energy transfer across a wavenumber cutoff $k_c$ between resolved scales ($k \leq k_c$) and subgrid scales ($k>k_c$).

It was shown by \citet{Domaradzki_2021a} that the task of modeling $T_{\textsc{sgs}}(k|k_c)$ can be productively split into finding the total SGS transfer/dissipation, integrated over $0<k\leq k_c$ and, separately, its distribution in wavenumbers $k$. The total SGS energy transfer across the cutoff $k_c$ is determined by the formula derived in \citep{Domaradzki_2021a,Domaradzki_2022b} 
\begin{equation}\label{eq:TSGS_total}
T_{\textsc{sgs}}(k_c)=\frac{1}{1-b}T_{\textsc{sgs}}^{res}(\frac{1}{2}k_c),
\end{equation}
where $T_{\textsc{sgs}}^{res}(\frac{1}{2}k_c)$ is the energy transfer computed for the resolved LES modes ($k \leq k_c$) and the cutoff $\frac{1}{2}k_c$, and $b$ is a constant that is a fraction of the total energy flux across $\frac{1}{2}k_c$ due to interactions with modes $k>k_c$.
Using multiple theoretical and DNS results for inertial range turbulence the rescaling factor $b$ was determined to be $b \approx 0.40$. 
The total resolved SGS energy transfer $T_{\textsc{sgs}}^{res}(\frac{1}{2}k_c)$ can be computed by integrating expression (\ref{eq:sgs}), written for cutoff $\frac{1}{2}k_c$, over all wavenumbers less than this cutoff.

\begin{figure}[t]
    \begin{center}
    \includegraphics[trim=30 30 0 0, clip, width=.6\columnwidth]{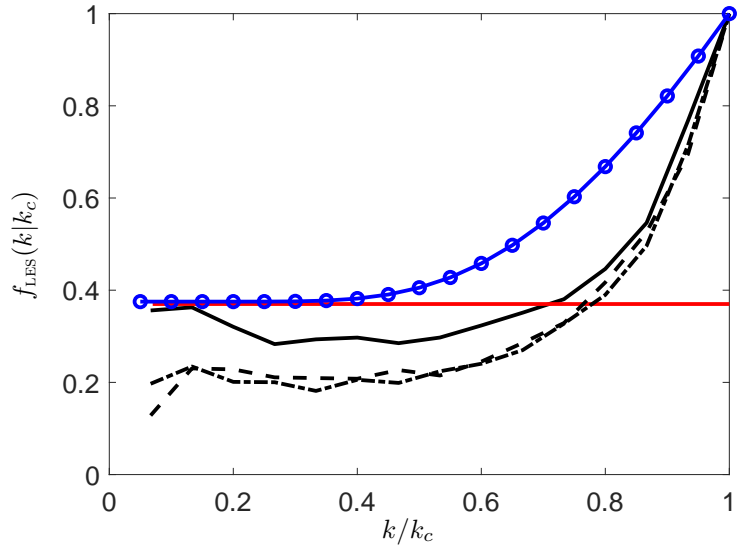}
    \put(-300,80){\rotatebox{90}{$f_{\textsc{les}}(k|k_c)$}}
    \put(-160,-10){$k/k_c$}
    \end{center}
    \caption{Spectral eddy viscosity shape functions. Solid line with symbols $\circ$: analytical theory of turbulence (EDQNM); horizontal solid line: asymptotic plateau value from the EDQNM theory. Remaining curves show shape functions computed from resolved LES fields for several cases in \citet{Domaradzki_2022a}.}
    \label{fig:SGSspectra}
\end{figure}

In practice, a wavenumber distribution of the resolved SGS energy transfer $T_{\textsc{sgs}}^{res}(k|\frac{1}{2}k_c)$ can be obtained from LES data during an actual run and cast in the form of the $k$-dependent eddy viscosity (\ref{eq:nu_th}), which is normalized to unity
$f^{res}_{\textsc{les}}(k|\frac{1}{2}k_c)=\nu^{res}_t(k|\frac{1}{2}k_c)/\nu^{res}_t(\frac{1}{2}k_c|\frac{1}{2}k_c)$. That last quantity, the eddy viscosity shape function, is rescaled from the test cutoff $(1/2)k_c$ to LES cutoff $k_c$, using the similarity variable $0 \leq k/k_{c} \leq 1$. Such computed eddy viscosities for several LES cases are shown in Fig. \ref{fig:SGSspectra}. Finally, the values of the eddy viscosity at low $k$ are modified to make them consistent with the asymptotic value provided by the analytical theories for the inertial range at $k/k_c \rightarrow 0$. Based on results from the EDQNM (eddy-damped quasi-normal Markovian) theory the plateau asymptotic value $p$ was determined as 0.37 of the peak value at the cusp, i.e., $p=0.37$ for the eddy viscosity shape function normalized to unity at $k_c$ (horizontal line in Fig. \ref{fig:SGSspectra}). The final shape function $f_{\textsc{les}}(k|k_c)$ has a similar shape as the analytical profile from the EDQNM theory, and comprise a constant plateau up to an intersection with a cusp of the resolved shape function $f^{res}_{\textsc{les}}(k|k_c)$, followed by the unmodified cusp part from the intersection point to $k=k_c$ that enhances small scale dissipations.

The details of shape function are used as a constraint to generate a desired distribution of SGS transfer, but more importantly, the total SGS dissipation estimated from equation (\ref{eq:TSGS_total}) needs to be addressed as well. Consequently, the prescription of the shape of interscale energy transfer in spectral space is implemented as follows. At each time step in simulations the eddy viscosity is
\begin{equation}\label{eq_nueddy}
\nu_t(k|k_c)=C_m f_{\textsc{les}}(k|k_c),
\end{equation}
where $C_m$ is an additional modeling constant and $f_{\textsc{les}}(k|k_c)$ is a shape function, determined as described above. Then $C_m$ is computed using known total SGS energy transfer as an integral constraint
\begin{equation}
T_{\textsc{sgs}}(k_c)=\int_{0}^{k_c}dk T_{\textsc{sgs}}(k|k_c)= -\int_{0}^{k_c}dk \; \nu_t(k|k_c)2 k^2 E(k),
\label{eq:int-constraint}
\end{equation}
which gives
\begin{equation}\label{eq_Cm}
C_m=\frac{-T_{\textsc{sgs}}(k_c)}{\int_{0}^{k_c}f_{\textsc{les}}(k|k_c) 2k^2 E(k)dk}.
\end{equation}
In LES runs the eddy viscosity (\ref{eq_nueddy}) is determined at each time step in simulations and used in the eddy viscosity term added to the Navier-Stokes spectral solver as a SGS modeling term.

The model is self-contained in the following sense. In equation (\ref{eq_Cm}) $T_{\textsc{sgs}}(k_c)$ is expressed in terms of SGS transfer among resolved scales $T_{\textsc{sgs}}^{res}(\frac{1}{2}k_c)$ (see Eq. (\ref{eq:TSGS_total})) computed at each time step in LES with the spectral eddy viscosity given by (\ref{eq_nueddy}). Similarly, the shape function $f_{\textsc{les}}(k|k_c)$ is computed at each time step from the resolved SGS energy transfer $T_{\textsc{sgs}}^{res}(k|\frac{1}{2}k_c)$, i.e., both factors in the formula (\ref{eq_nueddy}) are computed from information available in LES. In effect, the SGS model is not prescribed but obtained from the resolved SGS energy transfer $T_{\textsc{sgs}}^{res}(k|\frac{1}{2}k_c)$ in a given LES and well-established properties of the energy flux for the inertial range in the asymptotic limits $k/k_c \rightarrow 0$ and $k/k_c \rightarrow \infty$. Note also that since $T_{\textsc{sgs}}^{res}(k|\frac{1}{2}k_c)$ and $E(k)$ in general are time dependent, both factors in (\ref{eq_nueddy}) are also functions of time, $C_m(t)$ and $f_{\textsc{les}}(k,t|k_c)$.

The procedure described above was specifically designed, developed, and tested for isotropic, homogeneous turbulence at very high Reynolds numbers, simulated using pseudo-spectral Fourier methods. This may be considered of limited interest for LES of wall-bounded flows in complex geometries, wall modeling for high Reynolds/Mach number incompressible/compressible flows, and modeling additional physical phenomena such as heat transfer, chemical reactions, etc. However, this choice was a necessary first step in advancing the proposed method. Since isotropic turbulence is the case for which physical theories and DNS provide most complete and unambiguous information about energy transfer among scales of motion, this is also the case for which that information can be most directly exploited in model development. Additionally, it is not uncommon that new SGS models that show promise in LES of low Reynolds number turbulence fail catastrophically at high Reynolds numbers.
This is due to the fact that as the viscous dissipation in LES diminishes with an increasing Reynolds number, the model alone may fail to provide sufficient SGS dissipation, the similarity model of \citet{Bardina_1983} being a prominent example. After documenting the proper performance of the SGS methodology for isotropic turbulence at high Reynolds numbers, the current investigation proposes an extension of the methodology to physical space implementations, whose performance is assessed for wall-bounded flows.

\section{Physical space implementation}
\label{sec:nm}

In this section, the SGS modeling approach is extended to accommodate implementations in physical space. The primary goal is still to maintain the correct total SGS transfer based on information available in a LES run. Due to the small unresolved scales in LES and lack of asymptotic theory for inhomogeneous flows, it is not straightforward to estimate the total SGS dissipation directly. However, there are two useful observations from the modeling procedure applied to isotropic turbulence.

Firstly, as discussed in \citep{Domaradzki_2021a} the SGS transfer can also be computed using standard LES formulas for SGS dissipation in the physical space based on the SGS tensor and the resolved rate-of-strain tensor. Specifically, for general implementations in physical space, the kinetic energy transfer for a test-filtered field can be computed based on the subfilter scales (SFS)
\begin{equation}
  \tau_{ij}^{\textsc{sfs}} = \tau_{ij}^{\text{sim}} = \widetilde{\bar{u}_i \bar{u}_j} - \widetilde{\bar{u}}_i \widetilde{\bar{u}}_j,
  \label{eq:sim}
\end{equation}
where tilde $\widetilde{\enspace}$ denotes a test-filtering, whose characteristic filtering width $\widetilde{\Delta}$ is usually larger than the LES mesh size $\Delta$. Note that the current SFS stress is based on the resolved LES scales, which differs from the ``unresolved'' SFS studied in e.g. \citep{Rouhi2016}. The definition is analogous to the superscript ``$<$'' in equation (\ref{eq:SGS_en}) for $k \leq ak_c$ and $a<1$. Specifically, the SFS stress makes a clear distinction between the filter scale $\widetilde{\Delta}$ and the LES grid scale $\Delta$, the splitting of the former is performed only on resolved scales with either spectral or smooth filter. A detailed description of SFS stress and its applicability in LES is discussed in detail in the monograph of \citet{Berselli_2006}, among others. Note that the subfiltering stress is exactly the same as the well-known Bardina's similarity model \citep{Bardina_1980,Bardina_1983}. In general, the above tensor is multiplied by a constant coefficient $C_{\text{sim}}$, which is $O(1)$ based on $a$ $priori$ studies \citep{Liu_1994} and theoretical analysis \citep{Cook_1997}. Similarity-type models can achieve very high correlations in $a$ $priori$ analyses, but in $a$ $posteriori$ tests, due to the large backscatter, they are widely reported to be inadequately dissipative. In actual simulations with relatively low grid resolutions, it is normally necessary to apply $ad$ $hoc$ procedures to reduce the backscatter and stabilize the simulations (see e.g. \citep{Zang_1993,Liu_1994}). In \citet{Domaradzki_2021a} it was shown that SGS dissipation computed using (\ref{eq:sim}) and a Gaussian filtering can serve as an accurate proxy for the SGS dissipation computed using spectral methods with the spectral cutoff filter with $(1/2)k_c$ that is needed in expression (\ref{eq:TSGS_total}).

Another observation is that as long as the total SGS transfer was enforced, the details of its wavenumber distribution (i.e. its $k$-dependence) were of secondary importance. Specifically, in \citep{Domaradzki_2022a,Domaradzki_2022b} shape functions with different plateau levels and cusps were giving largely same results for energy spectra when the total transfer was computed using (\ref{eq:TSGS_total}) with the same parameter b=0.4. In particular, using a constant value of the shape function $f_{\textsc{les}}(k,t)=1$ provided accurate spectra in LES, except in the immediate vicinity of $k_c$ where the presence of the cusp was important. These both observations will be employed in the context of LES of the turbulent channel flow by using the SGS dissipation computed in the physical space using the similarity expression (\ref{eq:sim}) and converting the total SGS dissipation into the constant eddy viscosity model, corresponding to neglecting its $k$-dependence. 

In addition, from the perspective of kinetic energy transfer, whether the model takes a structural form or functional form is not crucial. Following the approach for isotropic turbulence, here we convert the model to the eddy viscosity form while maintaining the same total energy transfer. Recall that the similarity model can be considered as the SGS stress defined on the filtered grid. We restrict the estimation of eddy viscosity to filtered scales, i.e. the spectral support is $k \leq a k_c$ with $a < 1$. The kinetic energy transfer for the SFS stress based on the test-filtered field can be written in a similar way as the SGS dissipation
\begin{equation}
  \varepsilon_{\textsc{sfs}} = -\tau_{ij}^{\textsc{sfs}} \widetilde{\bar{S}}_{ij},
  \label{eq:epssfs}
\end{equation}
in which the filtered velocity field is used to compute the strain rate. Following the convention in LES, a positive value of $\varepsilon$ indicates energy dissipation. Note that when the subfilter stress or the similarity model is considered as a SGS model, it is multiplied by $\bar{S}_{ij}$ to compute the SGS dissipation. As discussed in \citep{Domaradzki_2021a}, due to the different spectral support for the SGS and SFS stresses, the latter cannot generate accurate SGS energy transfer. Specifically, the SF scales are mostly defined on $k \leq ak_c$, so the dissipation will be more dominant around $ak_c$ rather than the true LES cutoff $k_c$. The large backscatter at moderate wavenumbers near the filter cutoff is speculated to be a key reason to generate numerical instabilities and inaccurate predictions. To avoid the scale mismatch, we compute the total energy transfer only for the filtered scales, which can be used to obtain the effective eddy viscosity on LES grid
\begin{equation}
  \widetilde{\bar{\nu_t}}(\bm{x}) = \frac{\varepsilon_{\textsc{sfs}}(\bm{x})}{- |\bar{S}|^2(\bm{x})},
  \label{eq:nutf0}
\end{equation}
where $|\bar{S}| = \sqrt{2\bar{S}_{ij} \bar{S}_{ij}}$. Similar to the dynamic procedure \citep{Germano_1991}, in actual implementations, the resulting eddy viscosity (or the numerator and denominator, respectively) is spatially averaged to enhance numerical stability. For channel flow, it is natural to perform averaging over the homogeneous directions, for instance the planar averaging of coefficients in the dynamic Smagorinsky model \citep{Piomelli_1993}. In practice, we found that an averaging over one of the streamwise or spanwise direction is sufficient to make LES runs numerically stable and leads to slightly more accurate results than those from 2-D averaging. Consequently, all results shown in the current work are based on 1-D spanwise averaging, thus $\bar{\nu_t}$ depends only on $x$ and $z$ directions only.

Finally, to make the SGS dissipation in the present method equal to the ``true'' SGS dissipation from fully resolved simulations, a rescaling procedure based on the analyses in the previous section is necessary. As discussed before, the SFS dissipation defined on $k \leq ak_c$ only partially contributes to the total SGS dissipation. The subfilter scale dissipation at wavenumber $k \leq ak_c$ can be rescaled to the LES cutoff $k_c$ through a global rescaling factor $b$ (following the procedure in equation (\ref{eq:TSGS_total}))
\begin{equation}
  \varepsilon_{\textsc{sgs}} = \frac{1}{1-b} \varepsilon_{\textsc{sfs}}, \text{ or } \bar{\nu_t}(\bm{k}|k_c) = \frac{1}{1-b} \widetilde{\bar{\nu_t}}(\bm{k}|ak_c),
  \label{eq:nutk0}
\end{equation}
which allows us to recover the total SGS dissipation from only resolved scales. For high Reynolds number isotropic turbulence the parameter $b=0.4$ was determined solely from the asymptotic properties of the energy flux in \citep{Domaradzki_2022a,Domaradzki_2022b}. As discussed in \citet{Domaradzki_2021b} the value of $b$ decreases for test cutoff $(1/2)k_c$ in the dissipation range. For channel flow turbulence considered here this is very likely the case, as shown through the energy spectra that are steeper than the inertial range slope. Therefore for the same sharp spectral filter, we will use the lower limit for $b=0$, implying that the energy transfer is principally local and the total transfer is given by the resolved transfer, i.e., the nonlocal transfer between scales $k<(1/2)k_c$ and $k>k_c$ is negligible. The method will then be extended to smooth filters, making it possible to implement the model for fully inhomogeneous flows.

Following the notation in our previous works \citep{Anderson_2012,Domaradzki_2021a}, the resulting SGS model based on the concept of interscale energy transfer is still denoted as ``ITM''. The new model offers several clear advantages over traditional eddy viscosity models or similarity-type models. 
The present approach is established based on the connection between the SGS and SFS fields. It is well-known that the similarity model displays a very high correlation in $a$ $priori$ tests \citep{Bardina_1980,Liu_1994}. According to the recent analyses in \citet{Domaradzki_2021a,Domaradzki_2021b}, the spectral eddy viscosity obtained from test-filtered LES field is quantitatively close to the Chollet-Lesieur model \citep{Chollet_1981} based on the EDQNM theory and the widely-used spectral vanishing viscosity (SVV) formula \citep{Karamanos_2000,Lamballais_2011}. Therefore, compared with previous methods designed from the perspective of energy transfer (e.g. \citep{Anderson_2012,Cimarelli_2014}), constructing SGS dissipation based on the test-filtered velocity field has more solid theoretical foundations. Besides, the recasting to eddy viscosity in (\ref{eq:nutf0}) offers an additional constraint to the model, which can be regarded as a simple step to make the similarity model numerically more robust. Finally, a desired amount of total SGS dissipation is guaranteed through a simple but efficient rescaling procedure (\ref{eq:nutk0}), which accomplishes the fundamental requirement of functional modeling \citep{Sagaut_2006}. Employing the interscale energy transfer concept makes the model also more meaningful physically. The theoretical analyses and satisfactory simulation results for homogeneous, isotropic turbulence \citep{Domaradzki_2021a,Domaradzki_2021b} at high Reynolds numbers give us more confidence to extend the modeling strategy to more complex flows. As we will see in the $a$ $posteriori$ analysis in Sec. \ref{sec:res}, the ability of generating a proper amount of SGS dissipation among different scales is crucial to the success of an SGS model.

To perform LESs for turbulent channel flow, the incompressible Navier-Stokes equations (\ref{eq:ns}) are solved using a hybrid spectral/FDM (finite difference method) solver. Fourier pseudo-spectral method is used in the horizontal directions, dealiasing is performed with the 3/2 rule \citep{Canuto_2007}. The Chebyshev-Gauss-Lobatto points are used as the vertical grid, the derivatives are computed by a 4th order central difference scheme. A second order predictor-corrector scheme with semi-implicit Crank-Nicolson method for viscous term is used for time advancement, in which the time step $\Delta t$ is determined based on both the desired CFL number and the convergence in the corrector. A pressure-Poisson equation formulation is employed and an influence-matrix technique is applied \citep{Kleiser_1980}, which satisfies no-slip boundary conditions and the continuity constraint with machine accuracy. A constant mass flux for channel flow is enforced for all runs.

\section{Results}
\label{sec:res}

In this section, $a$ $posteriori$ analysis for fully developed turbulent channel flow at various Reynolds numbers is performed. In order to quantify the relative error in comparison with DNS benchmarks, we follow approach of \citet{Toosi_2017} with the overall error defined as
\begin{equation}
  Err_{\textsc{dns}} = \frac{1}{5} \frac{\int^b_a |U^+ - U^+_{\textsc{dns}}| d y^+}{\int^b_a U^+_{\textsc{dns}} d y^+} + \sum_{i,j} \frac{1}{5} \frac{\int^b_a |R^+_{ij} - R^+_{ij,\textsc{dns}}| d y^+}{\int^b_a \frac{1}{2} R^+_{kk,\textsc{dns}} d y^+},
  \label{eq:er}
\end{equation}
where $R_{ij}$ denotes the Reynolds stresses. The superscript ``$+$'' implies normalization in wall units, in which $y^+ = y u_\tau / \nu$, and $u_\tau$ is the frictional velocity. The coefficient $\frac{1}{5}$ is used to average the errors due to different quantities (only 4 non-vanishing Reynolds stresses for channel). The integrations are performed over half channel height $h$. As a reference, for our code the overall error ($Err_{\textsc{dns}}$) using equation (\ref{eq:er}) for $Re_\tau = u_\tau h/\nu \approx 590$ and the same resolution is less than 0.5\%, compared with data from \citet{Moser_1999}.

In addition, we will also show the relative error of the mean profile separately because the Reynolds stresses cannot be trusted if the mean velocity is already off. However, in some cases we noticed that the relative error computed from the first term on the r.h.s. of equation (\ref{eq:er}) is not very consistent with the observations from plots. This is likely due to the additional integration in the formula, which puts more weights on grid points with relatively larger mesh size. Ideally, we want the lower order statistics in LES to be almost the same as the DNS benchmark, so the value at each grid point should be equally important. Therefore, the following equation is used to compute the averaged relative error for mean flow
\begin{equation}
  Err_{\text{m}} = \frac{1}{N_y-2} \sum \frac{|U^+ - U^+_{\textsc{dns}}|}{U^+_{\textsc{dns}}},
  \label{eq:em}
\end{equation}
where $N_y$ is the number of grid points in the vertical direction. For channel flow, $U^+$ at the walls vanish, so the wall nodes are excluded and $N_y-2$ points are used for averaging.

\begin{table}[ht]
  \centering
  \caption{\label{tab:cases} Grid resolution ($x,y,z$, respectively) for current LES and benchmark DNS. }
  \begin{tabular}{c|c|c|c|c}
  \hline
  \hline
  Case& $Re_{\tau}$ & Domain     & Resolution   & $N/N_{DNS} (\%)$   \\\hline
  LES & 180 & $4\pi \times 2 \times 2\pi$   & $32 \times 65 \times 32$     & 3.1492\% \\
  DNS & 180 & $4\pi \times 2 \times \frac{4}{3}\pi$ & $128 \times 129 \times 128$ & \rule[0.5ex]{1em}{0.55pt} \\
  \hline
  LES & 1000 & $2\pi \times 2 \times \pi$   & $60 \times 150 \times 60$   & 0.0225\% \\
  DNS & 1000 & $8\pi \times 2 \times 3\pi$   & $2304 \times 512 \times 2048$& \rule[0.5ex]{1em}{0.55pt} \\
  \hline
  LES & 2000 & $2\pi \times 2 \times \pi$   & $60 \times 200 \times 60$   & 0.0075\% \\
  DNS & 2000 & $8\pi \times 2 \times 3\pi$   & $4096 \times 768 \times 3072$& \rule[0.5ex]{1em}{0.55pt} \\
  \hline
  \hline
  \end{tabular}
\end{table}

The Reynolds number and corresponding resolutions for all simulations are summarized in Table.~\ref{tab:cases}. All results were gathered within the statistically steady state (when a linear profile of Reynolds shear stress budget is observed \citep{Vinuesa_2016}) and compared with the DNS benchmark from \citet{Moser_1999} and \citet{Lee_2015} after adequate time averaging.

\subsection{Recasting of similarity model based on sharp spectral filter}
\label{sec:sim}

Following the previous works on homogeneous isotropic turbulence \citep{Domaradzki_2021a,Domaradzki_2021b}, the extended interscale energy transfer model (ITM) is first applied to the test-filtered velocity based on the sharp spectral cutoff filter. For channel flow, a cutoff at $\frac{1}{2} k_c$ is applied to the horizontal directions. Note that the similarity model (\ref{eq:sim}) generates an inhomogeneous SGS stress in the vertical direction, so the modeling is still three-dimensional. The model is then converted to the eddy viscosity form following the procedures in equations (\ref{eq:epssfs})$\sim$(\ref{eq:nutk0}).

For inhomogeneous flow defined in physical space, it is not straightforward to evaluate the rescaling factor $b$ theoretically. As explained in the discussion following equation (\ref{eq:nutk0}), due to the overall steeper channel flow energy spectra than the -5/3 scaling in the inertial range, the optimal values of $b$ are likely smaller than those for HIT . In addition, we seek reference from well-established results. For classic eddy viscosity models such as the Smagorinsky model \citep{Smagorinsky_1963}, the recommended modeling coefficient (i.e. the Smagorinsky coefficient $C_s$) for channel flow is normally chosen to be 0.1 or smaller \citep{Piomelli_1993}, lower than the theoretically predicted value $C_s \approx 0.16$ for homogeneous turbulence derived by \citet{Lilly_1967}. Eventually $C_s^2$ is used to compute the eddy viscosity in the model. Recall that a rescaling factor $b \approx 0.4$ for test-filtered field $k<\frac{1}{2}k_c$ is used in our previous analysis of HIT, implying a multiplication of $5/3$ of the resulting eddy viscosity, which likely needs to be reduced based on the experience of Smagorinsky model. As a first attempt, we choose the lower limit $b=0$ for all cases in this subsection.

\begin{table}[ht]
  \centering
  \caption{\label{tab:ct} Results for ITM performed with $\frac{1}{2} k_c$ spectral cutoff filter and $b=0$. }
  \begin{tabular}{c|c|c|c|c|c}
  \hline
  \hline
$Re_\tau \approx 180$                   & UDNS     & static Smag. & dynamic Smag. & ITM     & ITM, remove $\varepsilon_{\nu}$ \\
\hline
$Err_{\textsc{dns}}$ (\%)               & 5.3229   & 4.9990     & 3.7915       & 4.9463   & 3.8501                           \\
$Err_{\text{m}}$ (\%)                   & 2.6682   & 2.2988     & 2.4149       & 0.9207   & 1.2736                           \\
$\varepsilon_{\textsc{sgs}}$           & 0.00E+00 & 1.37E-04   & 1.24E-04     & 1.59E-04 & 1.47E-04                         \\
$T_\nu$                                 & 1.16E-03 & 9.77E-04   & 9.75E-04     & 9.85E-04 & 9.88E-04                         \\
$\varepsilon_{\textsc{sgs}}/T_\nu$ (\%) & 0.0000   & 14.0127     & 12.7563       & 16.1303 & 14.8808                         \\
\hline
$Re_\tau \approx 1000$                 & UDNS     & static Smag. & dynamic Smag. & ITM     & ITM, remove $\varepsilon_{\nu}$ \\
\hline
$Err_{\textsc{dns}}$ (\%)               & 9.9499   & 1.9066     & 1.9424       & 2.1379   & 2.5271                           \\
$Err_{\text{m}}$ (\%)                   & 6.0527   & 1.9438     & 0.3400       & 1.2878   & 1.4458                           \\
$\varepsilon_{\textsc{sgs}}$           & 0.00E+00 & 1.61E-04   & 1.67E-04     & 1.77E-04 & 1.69E-04                         \\
$T_\nu$                                 & 7.10E-04 & 5.44E-04   & 5.28E-04     & 4.99E-04 & 5.21E-04                         \\
$\varepsilon_{\textsc{sgs}}/T_\nu$ (\%) & 0.0000   & 29.6133     & 31.6650       & 35.4121 & 32.3488                         \\
\hline
$Re_\tau \approx 2000$                 & UDNS     & static Smag. & dynamic Smag. & ITM     & ITM, remove $\varepsilon_{\nu}$ \\
\hline
$Err_{\textsc{dns}}$ (\%)               & 8.0176   & 3.7718     & 2.8796       & 2.7306   & 4.0711                           \\
$Err_{\text{m}}$ (\%)                   & 7.1752   & 1.5276     & 3.6931       & 1.5049   & 0.4633                           \\
$\varepsilon_{\textsc{sgs}}$           & 0.00E+00 & 2.35E-04   & 1.89E-04     & 1.98E-04 & 1.88E-04                         \\
$T_\nu$                                 & 3.94E-04 & 3.78E-04   & 3.84E-04     & 3.74E-04 & 3.99E-04                         \\
$\varepsilon_{\textsc{sgs}}/T_\nu$ (\%) & 0.0000   & 62.2488     & 49.1604       & 52.8578 & 47.1868                         \\
\hline
  \hline
  \end{tabular}
\end{table}

\begin{figure}[t]
\subfloat
{%
  \includegraphics[width=0.49\columnwidth]{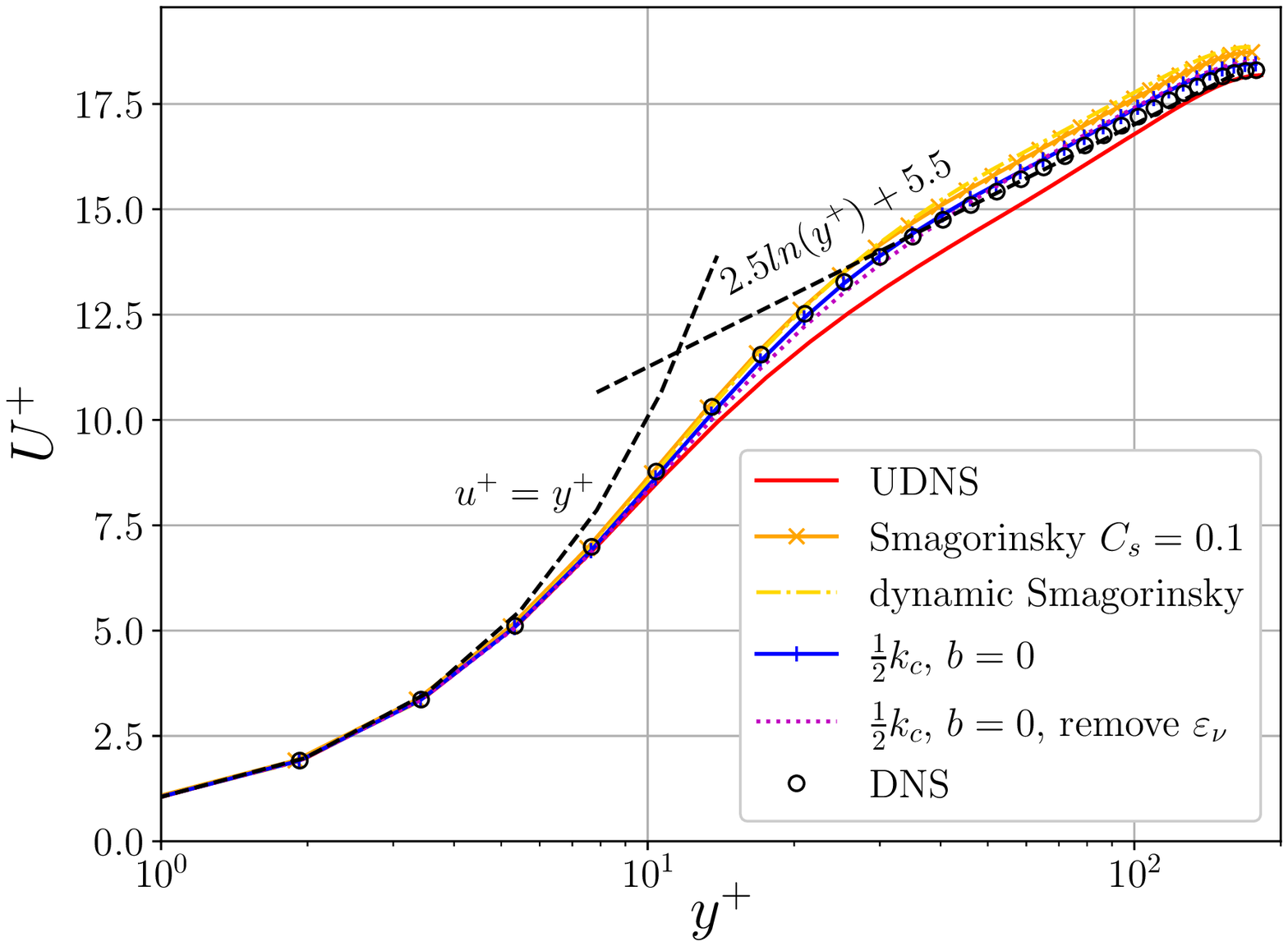}%
}
\subfloat
{%
  \includegraphics[width=0.49\columnwidth]{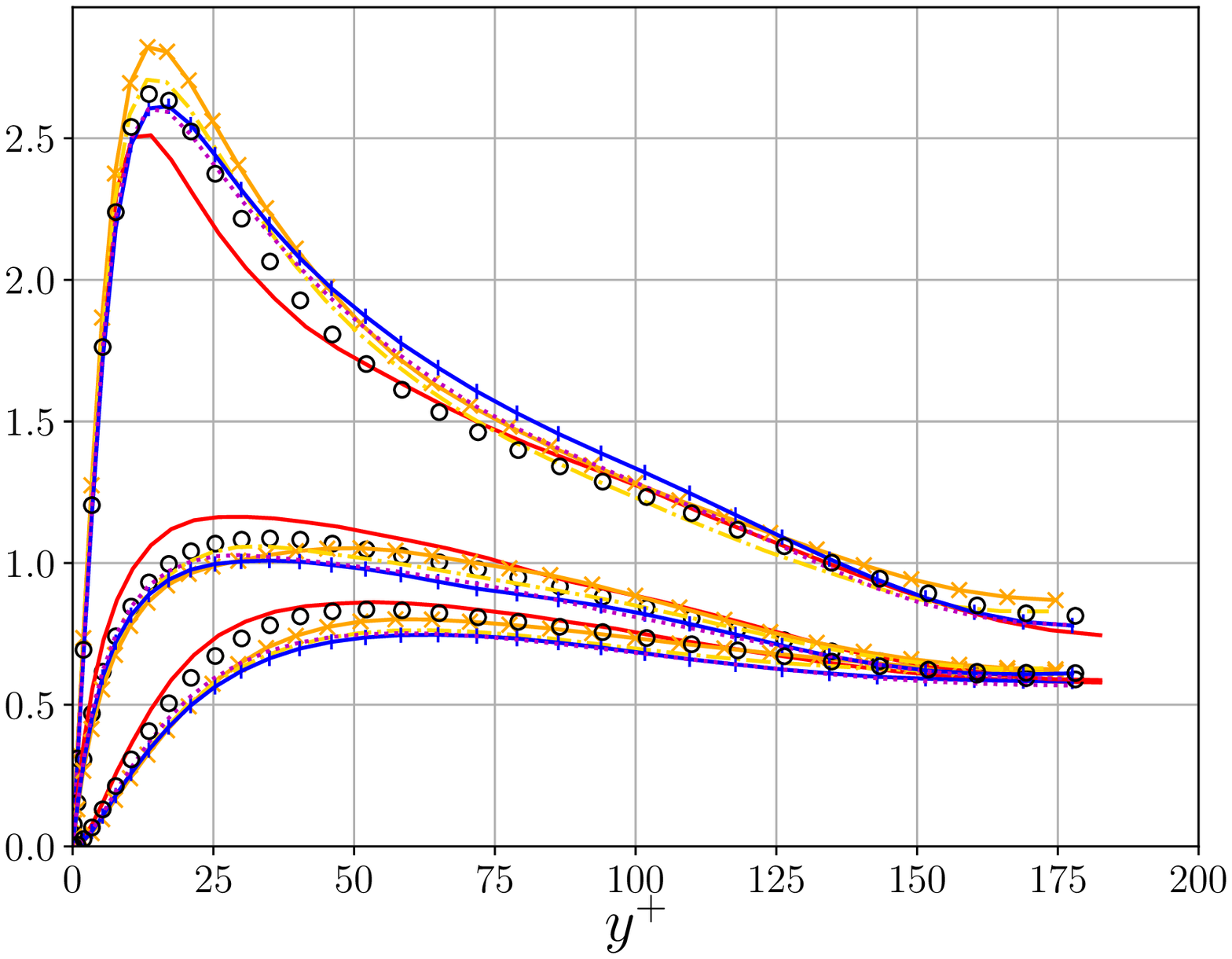}%
}
\caption{Mean velocity profiles (left) and rms velocities (right) for $Re_\tau=180$, the ITM is performed with $\frac{1}{2} k_c$ spectral cutoff filter and $b=0$.}
\label{fig:Re180ct}
\end{figure}

\begin{figure}[t]
\subfloat
{%
  \includegraphics[width=0.49\columnwidth]{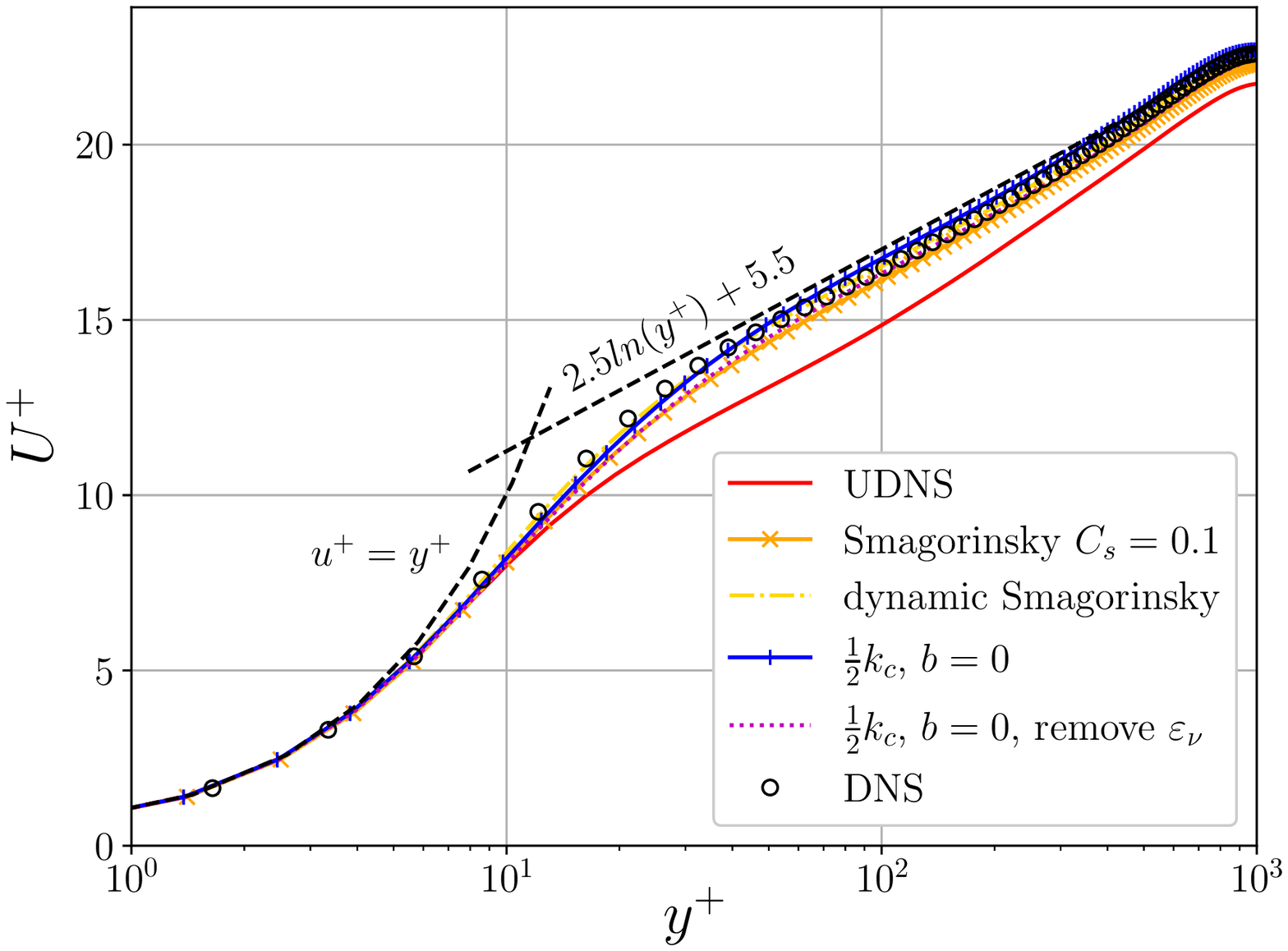}%
}
\subfloat
{%
  \includegraphics[width=0.49\columnwidth]{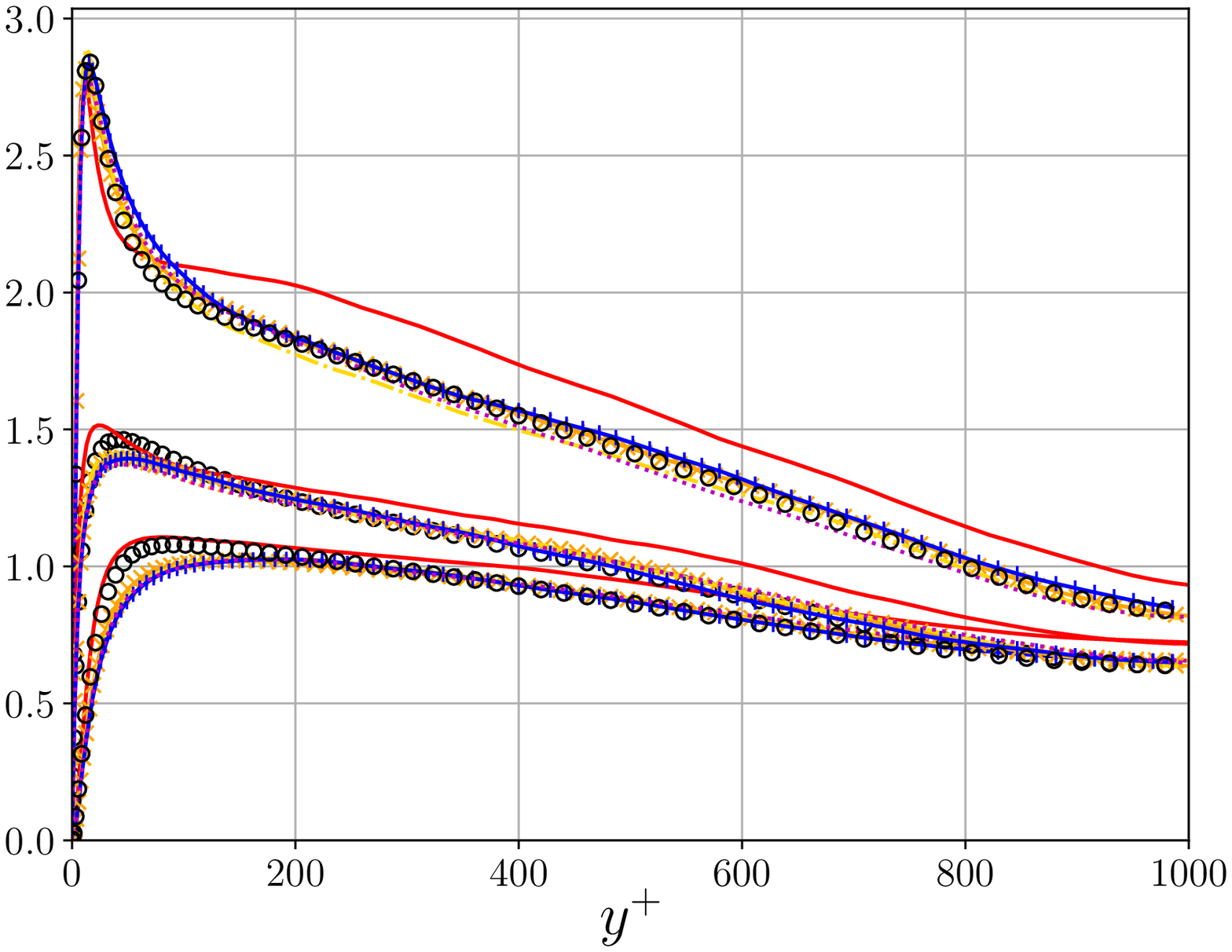}%
}
\caption{Mean velocity profiles (left) and rms velocities (right) for $Re_\tau=1000$, the ITM is performed with $\frac{1}{2} k_c$ spectral cutoff filter and $b=0$.}
\label{fig:Re1000ct}
\end{figure}

\begin{figure}[t]
\subfloat
{%
  \includegraphics[width=0.49\columnwidth]{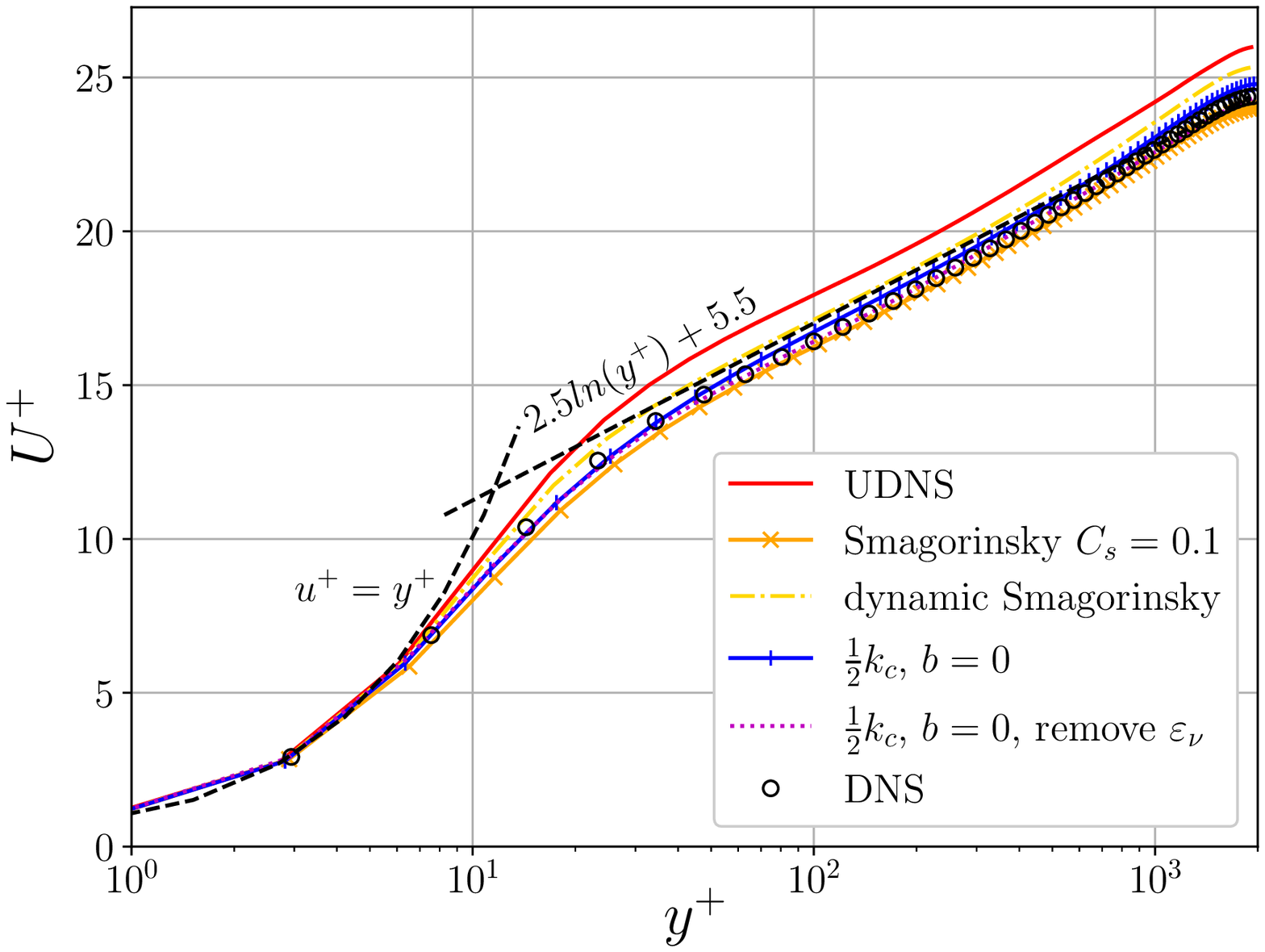}%
}
\subfloat
{%
  \includegraphics[width=0.49\columnwidth]{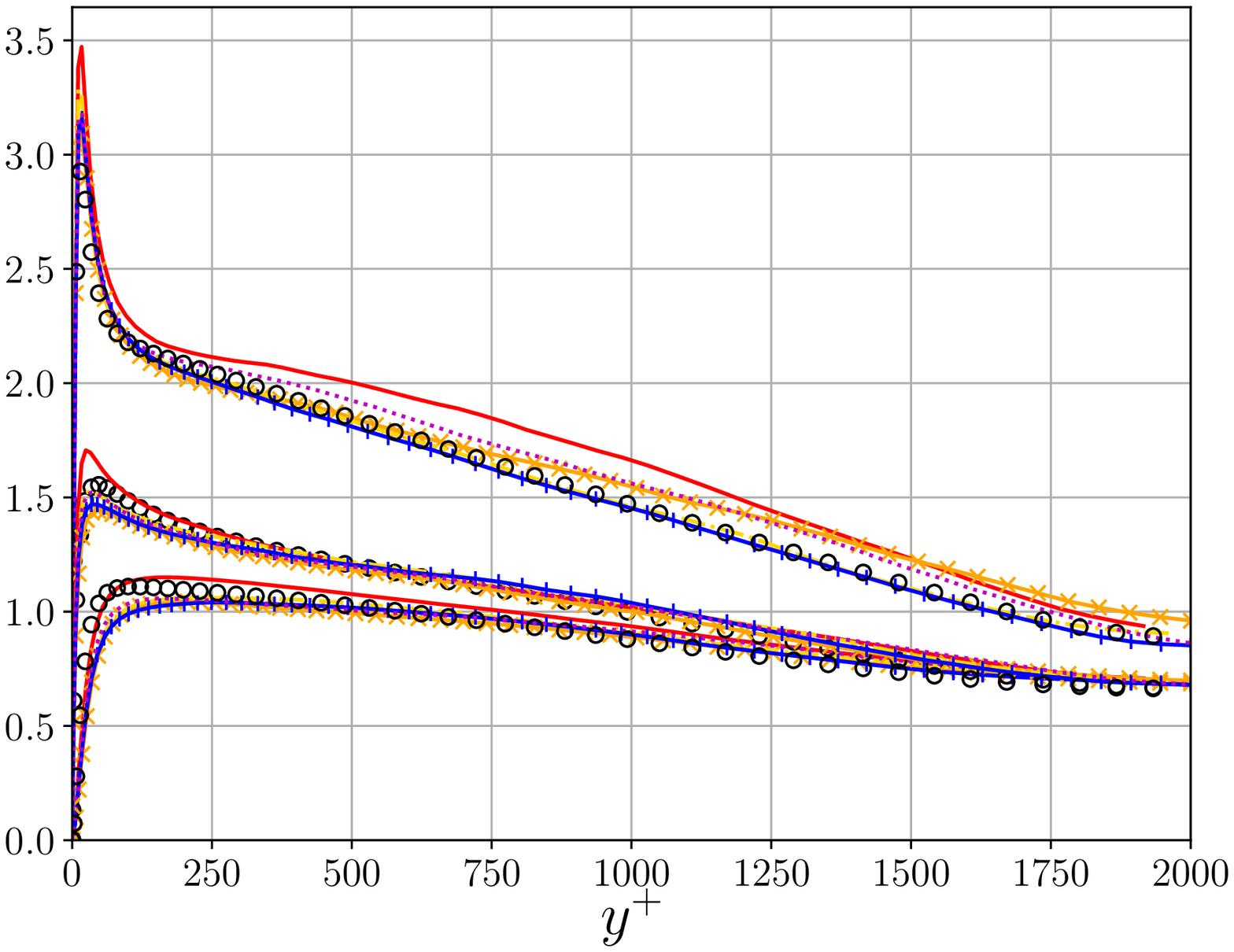}%
}
\caption{Mean velocity profiles (left) and rms velocities (right) for $Re_\tau=2000$, the ITM is performed with $\frac{1}{2} k_c$ spectral cutoff filter and $b=0$.}
\label{fig:Re2000ct}
\end{figure}

For various Reynolds numbers, the performance of our new model is compared with the no model simulations and LES with the Smagorinsky models in Figs \ref{fig:Re180ct}$\sim$\ref{fig:Re2000ct}. Corresponding results are also summarized in Table \ref{tab:ct}. Specifically, ``UDNS'' implies the no-model under-resolved DNS runs. The static Smagorinsky denoted as ``static Smag.'' always adopts a common choice of $C_s=0.1$ and the van Driest damping function \citep{Van_Driest_1956}. The dynamic Smagorinsky model based on the Germano identity \citep{Germano_1991} and extended in \citep{Lilly_1992} is used to compute the modeling coefficients is denoted as ``dynamic Smag.'', which applies a local averaging and zero-clipping of $C_s$ as described our previous works \citep{Cadieux_2015,Sun_2018}. The SGS dissipation and the viscous transfer $T_\nu = -\bm{u} \nabla^2 \bm{u}$ in the table were averaged over the entire channel. For higher Reynolds numbers, the contribution of SGS dissipation for all models is large, indicating the significant under-resolution in the context of DNS. For relatively low resolutions used in the present study (see also Table \ref{tab:cases}), no-model UDNS runs are incapable of generating satisfactory results. The Smagorinsky models can reproduce low order statistics accurately for some Reynolds numbers, but in general the accuracy is not as good as for the interscale model (the second column from the right). In particular, our new model focuses on providing an appropriate amount of total SGS dissipation, so that the mean velocity profiles are reproduced more faithfully. On the other hand, unlike our previous work for HIT in which the spectral eddy viscosity can be directly computed, in the physical space implementation less attention is paid to generate a desired distribution of energy transfer in spectral space, so the overall errors $Err_{\textsc{dns}}$ does not show a clear advantage compared to the Smagorinsky models. Nevertheless, the main objective of the current work is not to find an optimal modeling coefficient for various Reynolds numbers and the given turbulence flow; it is enough to see that a fixed rescaling factor $b=0$ suffices to generate satisfactory results for all cases.

\begin{figure}[t]
\subfloat
{%
  \includegraphics[width=0.49\columnwidth]{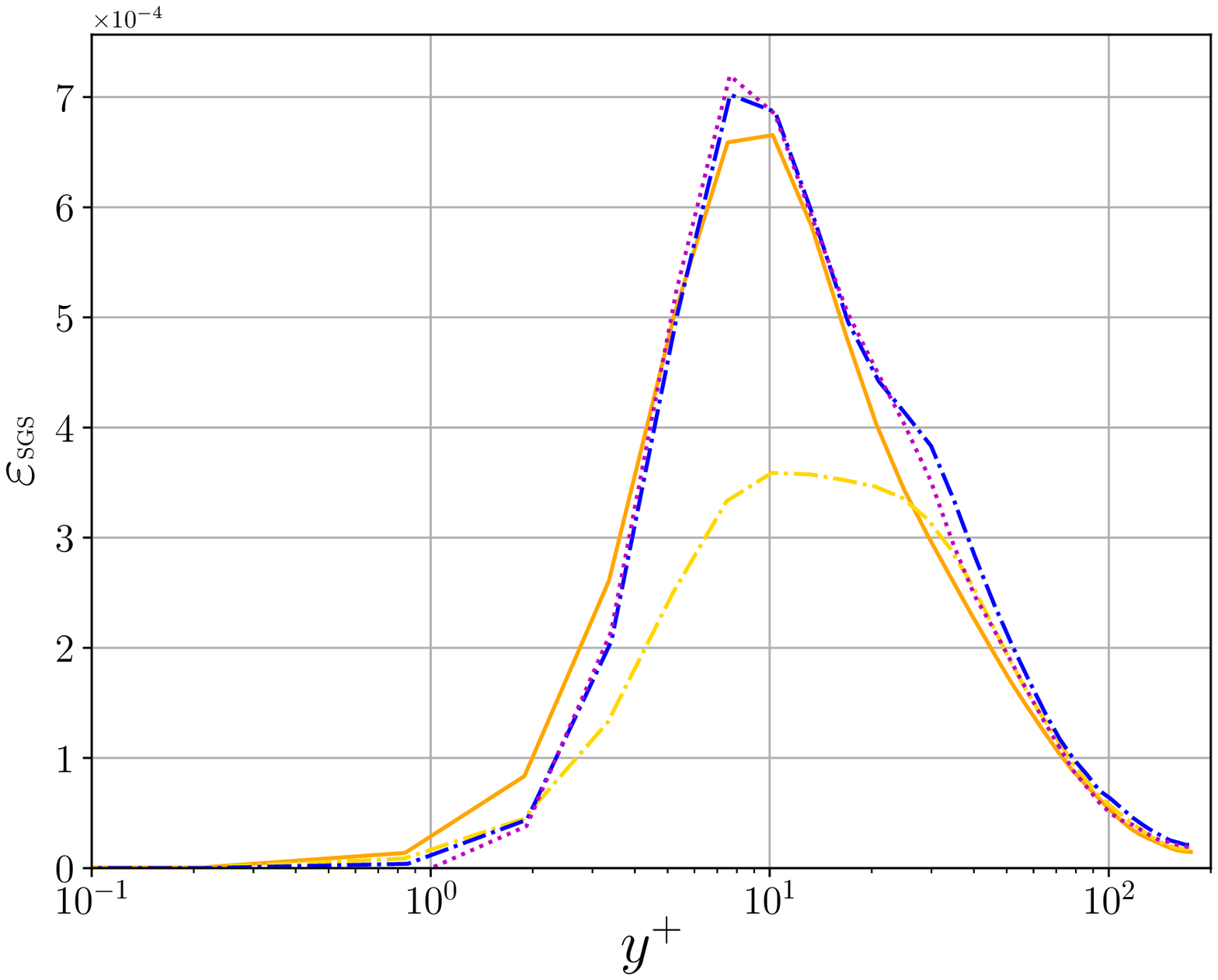}%
}
\subfloat
{%
  \includegraphics[width=0.49\columnwidth]{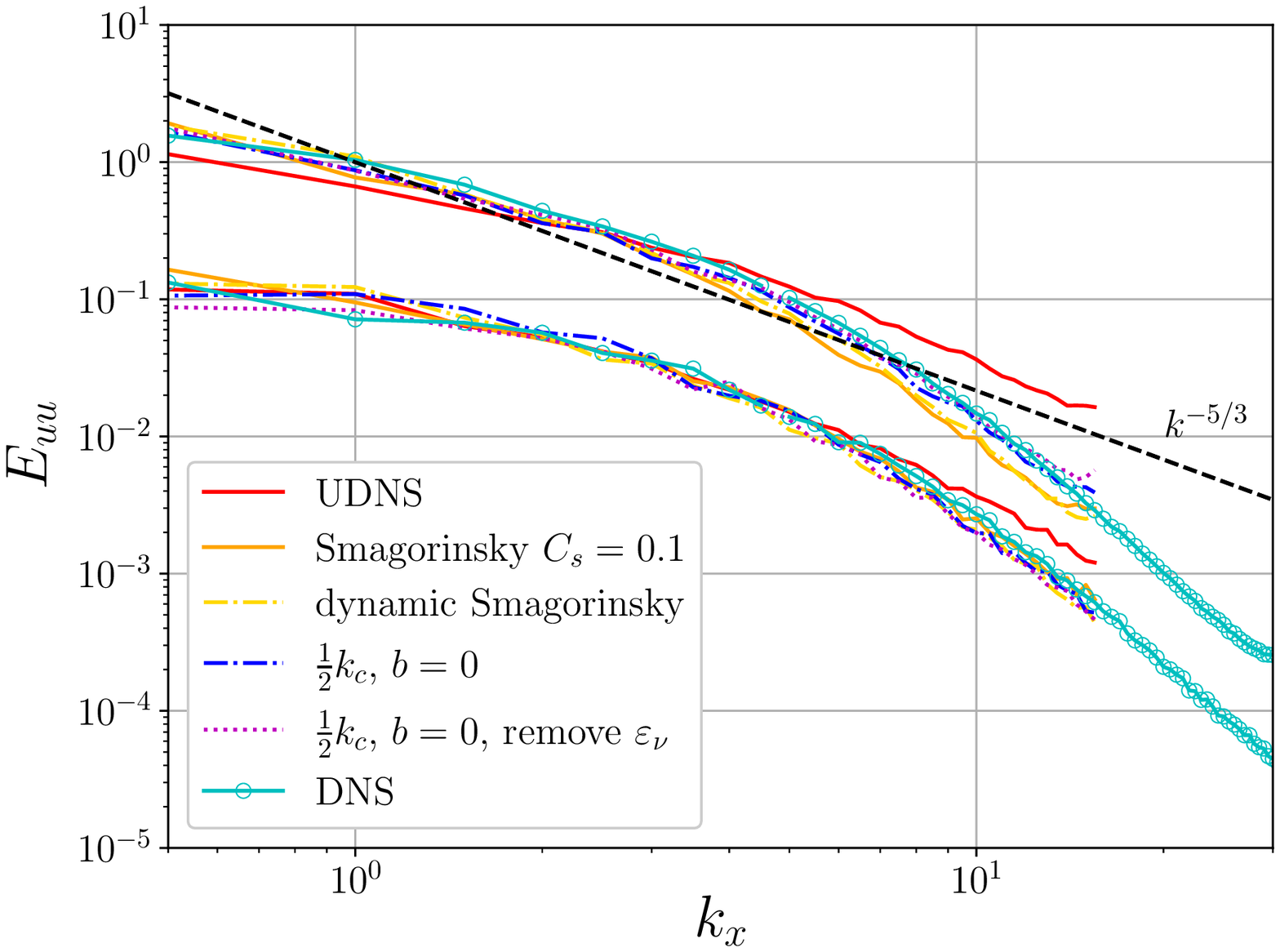}%
}
\caption{Horizontally averaged SGS dissipation (left) and 1-D streamwise energy spectra (right) for $Re_\tau=180$. The spectra in the upper part of the figure are for locations inside the buffer layer ($y^+ \approx 20$), and the spectra in the lower part are for locations close to the center ($y^+ \approx 180$). The ITM is performed with $\frac{1}{2} k_c$ spectral cutoff filter and $b=0$.}
\label{fig:Re180ct1}
\end{figure}

\begin{figure}[t]
\subfloat
{%
  \includegraphics[width=0.49\columnwidth]{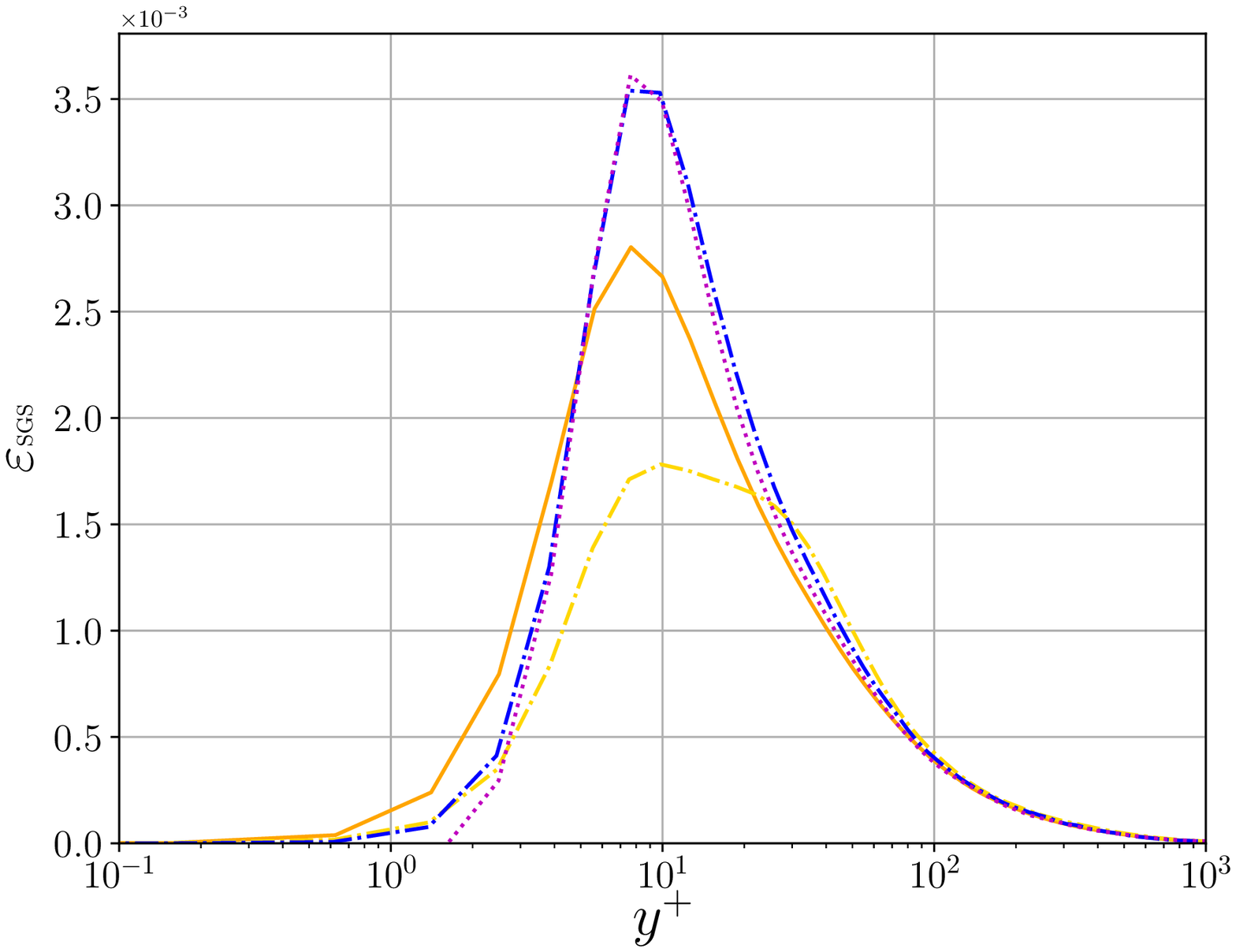}%
}
\subfloat
{%
  \includegraphics[width=0.49\columnwidth]{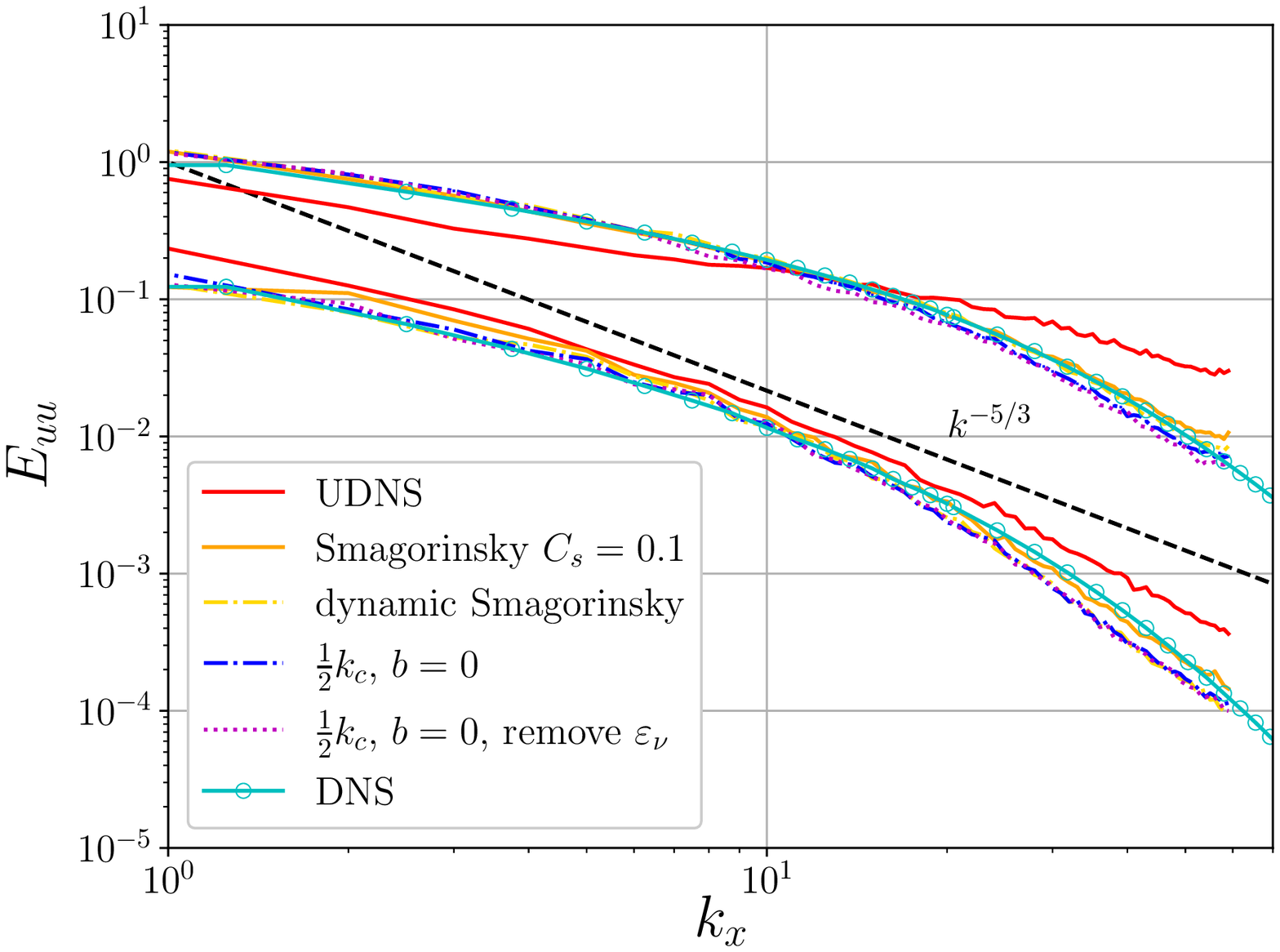}%
}
\caption{Horizontally averaged SGS dissipation (left) and 1-D streamwise energy spectra (right) for $Re_\tau=1000$. The spectra in the upper part of the figure are for locations inside the buffer layer ($y^+ \approx 20$), and the spectra in the lower part are for locations close to the center ($y^+ \approx 1000$). The ITM is performed with $\frac{1}{2} k_c$ spectral cutoff filter and $b=0$.}
\label{fig:Re1000ct1}
\end{figure}

\begin{figure}[t]
\subfloat
{%
  \includegraphics[width=0.49\columnwidth]{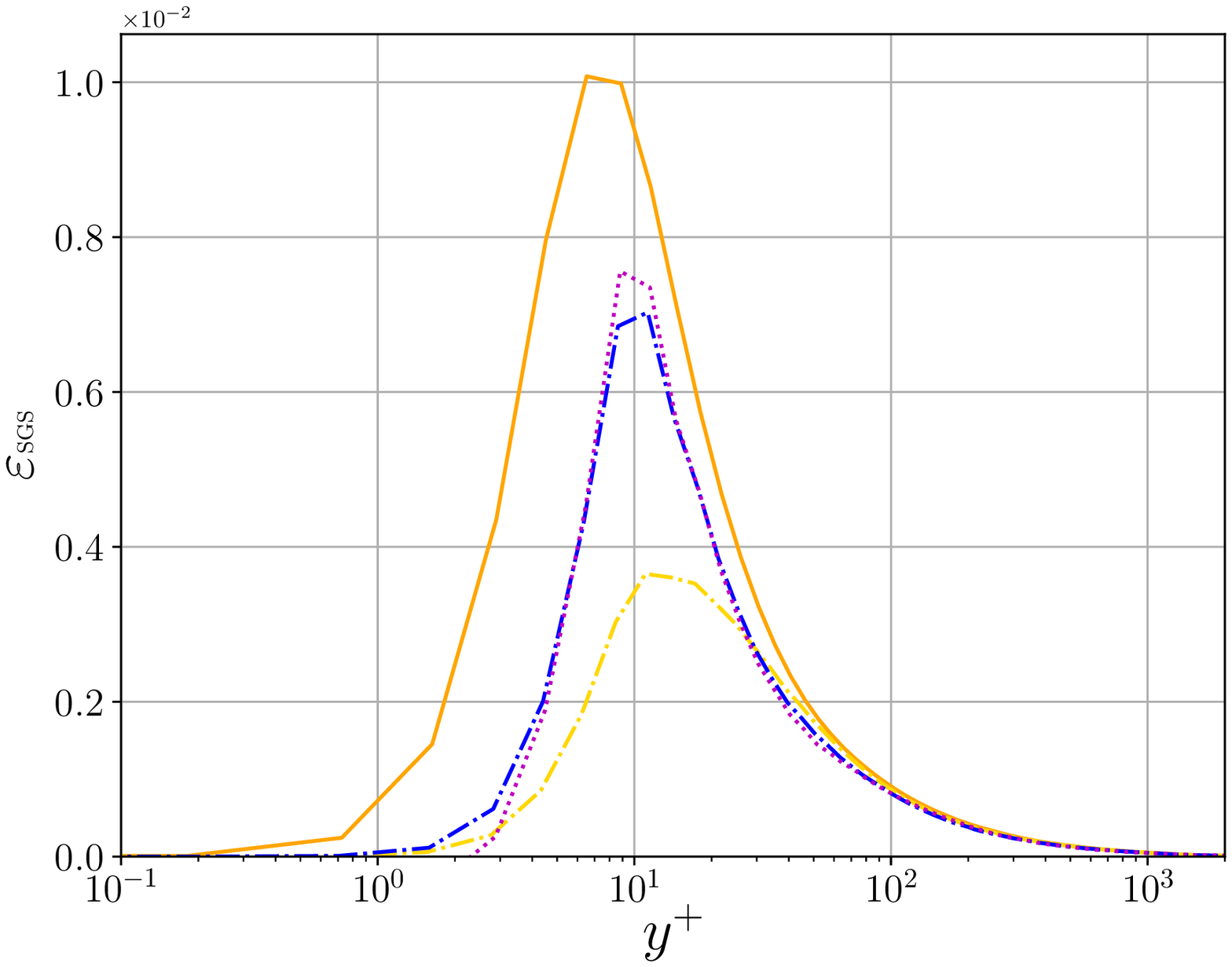}%
}
\subfloat
{%
  \includegraphics[width=0.49\columnwidth]{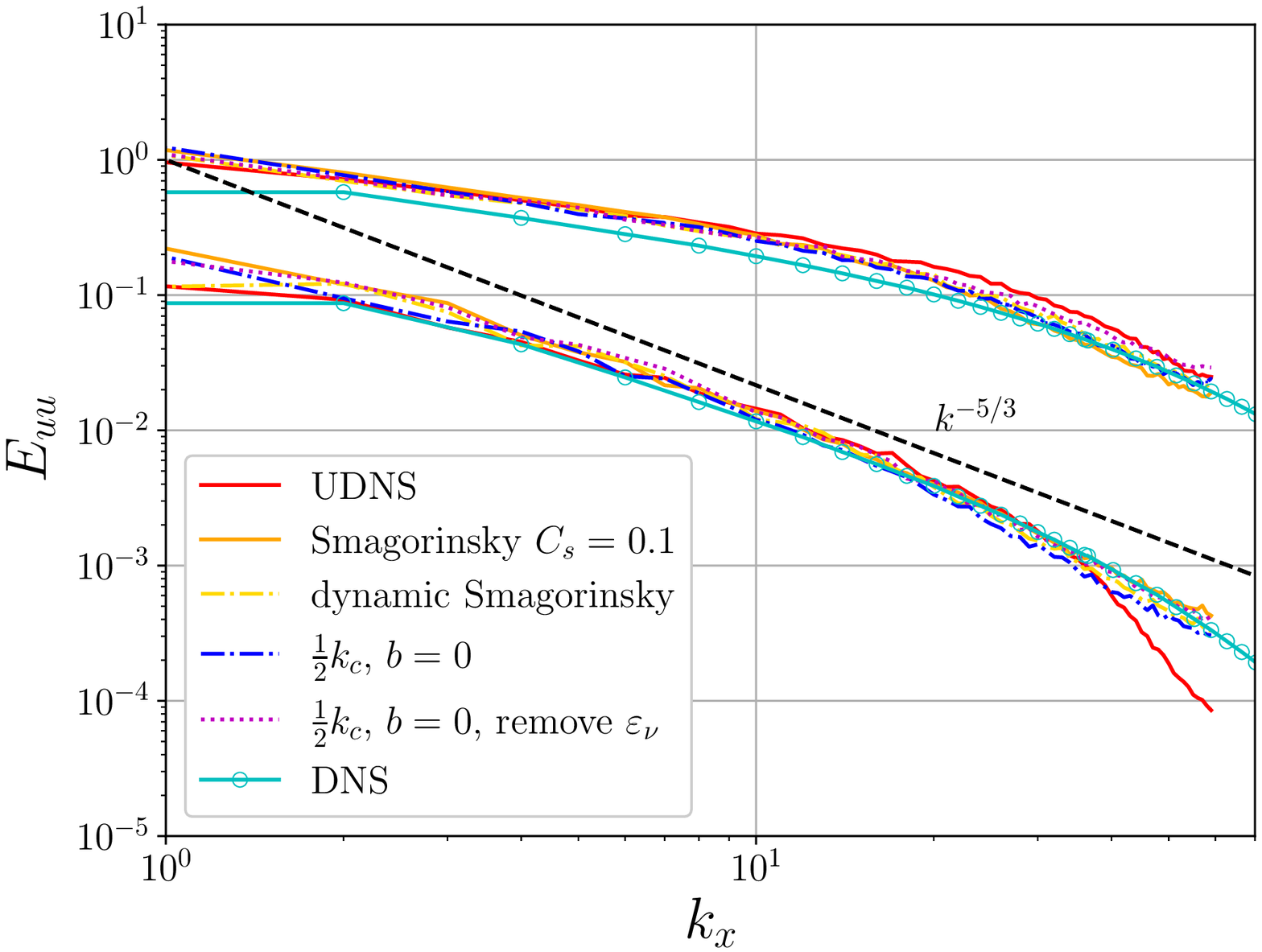}%
}
\caption{Horizontally averaged SGS dissipation (left) and 1-D streamwise energy spectra (right) for $Re_\tau=2000$. The spectra in the upper part of the figure are for locations inside the buffer layer ($y^+ \approx 20$), and the spectra in the lower part are for locations close to the center ($y^+ \approx 2000$). The ITM is performed with $\frac{1}{2} k_c$ spectral cutoff filter and $b=0$.}
\label{fig:Re2000ct1}
\end{figure}

The energy dissipation from the present model is further evaluated from the SGS dissipation and 1-D energy spectra in Figs \ref{fig:Re180ct1}$\sim$\ref{fig:Re2000ct1}. The vertical distribution of time averaged SGS dissipations are plotted in log scale to show the differences more clearly, which might be slightly misleading, as explained below. Based on results in Table \ref{tab:ct}, overall the globally averaged SGS dissipation of the new model with a fixed $b$ generated similar amount of SGS dissipation as the dynamic Smagorinsky model, especially the higher $Re$'s. The peak of $\varepsilon_{\textsc{sgs}}$ is mostly inside the buffer layer where the production of turbulent kinetic energy is the largest, and which is much higher for the new model. However, for relatively large $Re$'s the buffer layer constitutes only a small fraction of the channel height. The dynamic Smagorinsky model predicts a relatively more uniform distribution of $\varepsilon_{\textsc{sgs}}$ across the channel, leading to a comparable amount of total SGS dissipation as the ITM after vertical integration. As a consequence, for the dynamic Smagorinsky model in some cases a small amount of energy accumulation near the LES cutoff $k_c$ can be observed from the energy spectra inside the buffer layer. On the other hand, for all simulations with the new model the energy spectra follow the DNS benchmark well and nonphysical energy accumulation never occurs, indicating that the present model suffices to regularize the energy cascade for the resolved scales.

It is noteworthy that even though the volume-averaged SGS dissipations between the ITM and the dynamic Smagorinsky model are comparable, the primary objective for the dynamic procedure \citep{Germano_1991} and the interscale energy transfer concept is different. For a given SGS model, the dynamic procedure attempts to find the best estimates of modeling coefficients, while the interscale energy transfer model focuses on reproducing the correct amount of total SGS transfer for resolved scales. It is interesting to see that even though $b=0$ is used, implying the SFS dissipation computed from resolved scales is approximately equal to the total SGS dissipation, the averaged dissipations from our new approach are still larger than those from the dynamic Smagorinsky model. According to \citet{Domaradzki_2021b}, the rescaling factor is estimated based on properties of energy flux in the infinite inertial range, in particular, on the assumption of the constant energy flux across the spectrum. However, this assumption does not hold for channel flow in which the inertial range is not present. To improve the prediction of the total SGS dissipation we should account for the fact that the energy flux at the LES cutoff $k_c$ is equal to the energy flux at the test cutoff $(1/2)k_c$ decreased by the viscous dissipation in the range between these cutoffs
\begin{equation}
 \varepsilon_{\textsc{sgs}} \approx \varepsilon_{\textsc{sfs}} - \left[ \varepsilon_\nu\left(k \leq k_c \right) - \varepsilon_\nu\left(k \leq \frac{1}{2} k_c \right)\right],
 \label{eq:sgsmodified}
\end{equation}
where $\varepsilon_\nu$ denotes the viscous dissipation computed for indicated wavenumbers.

The model based on the modified total SGS transfer (\ref{eq:sgsmodified}) is denoted as ``remove $\varepsilon_\nu$'' in Table \ref{tab:ct} and Figs \ref{fig:Re180ct}$\sim$\ref{fig:Re2000ct}. The SGS dissipations slightly decreased compared with the original model and are very close to corresponding values obtained for the dynamic Smagorinsky model. Overall the relative errors before and after the modification do not differ by much. This is because while the total SGS dissipation becomes smaller in the modified procedure, it mostly reduces the small scale dissipation in the second half of wavenumbers. For the low order statistics of interest in LES, the energy at small scales near the cutoff wavenumber is normally less important. In previous simulations of homogeneous isotropic turbulence with ITM \citep{Domaradzki_2021a,Domaradzki_2021b}, a cusp with significantly higher spectral eddy viscosity is generated near $k_c$. In fact, an enhancement of energy dissipation specifically for relatively small scales is widely used in different types of LES modeling, such as the SVV \citep{Maday_1989,Karamanos_2000}, hyperviscosities \citep{Lamballais_2011,Lamballais_2021} and low-pass solution filtering \citep{Tantikul_2010,Sun_2018}. Therefore, the original ITM without removing $\varepsilon_\nu$ is more recommended as a general approach. Nevertheless, in the current work both methods were able to provide sufficient total dissipation and avoid energy accumulation at the LES cutoff $k_c$, suggesting the current procedure is general and not very sensitive to specific implementations.

\begin{figure}[t]
\subfloat
{%
  \centering
  \includegraphics[trim=0 0 150 0, clip, width=1.\columnwidth]{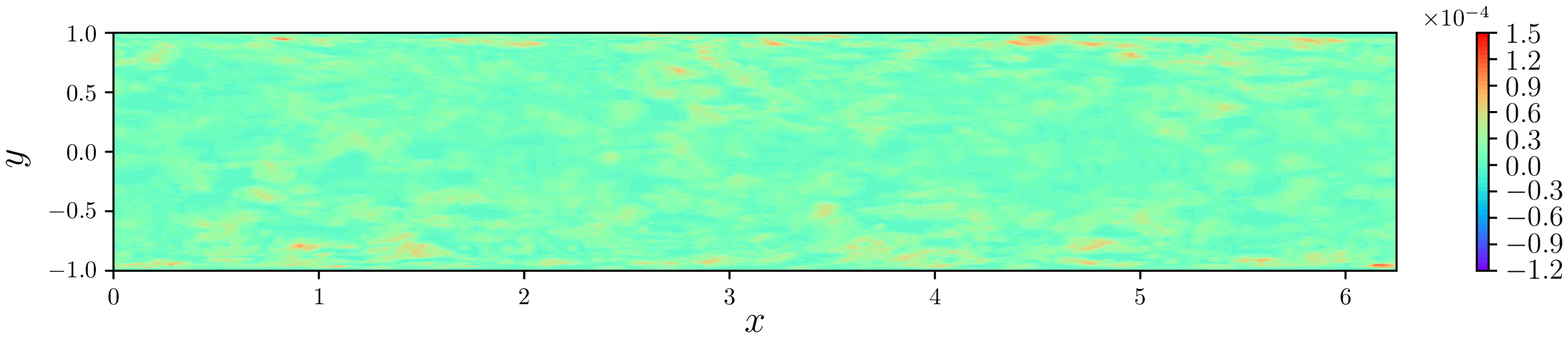}%
}
\put(-440,75){{\large $(a)$}}
\hfill
\subfloat
{%
  \centering
  \includegraphics[trim=60 0 210 0, clip, width=0.84\columnwidth]{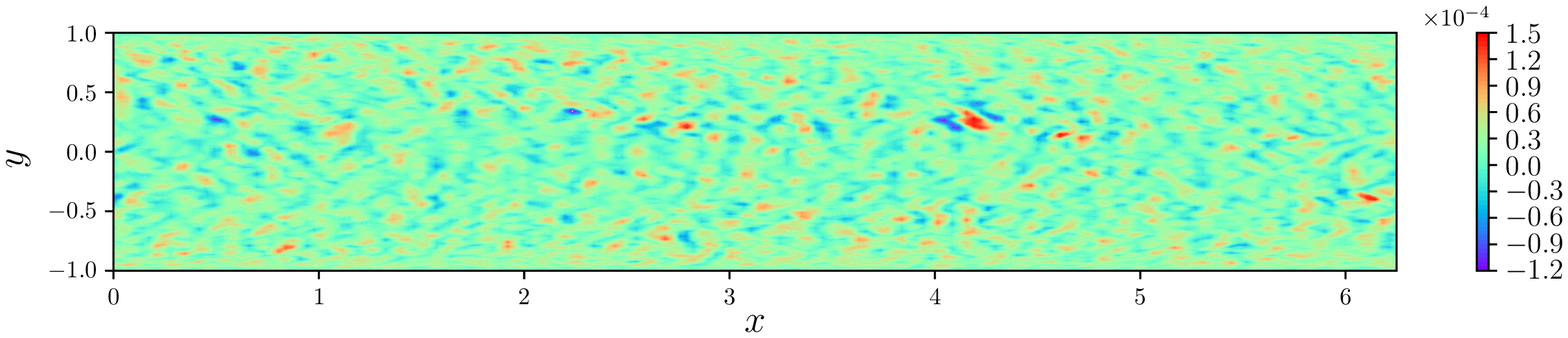}%
}
\put(-405,75){{\large $(b)$}}
\caption{Instantaneous eddy viscosity at $Re_\tau \approx 1000$, $(a)$ the dynamic Smagorinsky model, $(b)$ ITM computed from sharp spectral filter and $b=0$.}
\label{fig:Re1000nut}
\end{figure}

Although the total SGS dissipations for the models in the last 3 columns of Table \ref{tab:ct} are similar, we noticed that the new interscale method is overall more reliable for different $Re$ and resolutions. We attribute the differences to the fact that the Smagorinsky models implemented in the present work are purely dissipative, i.e. the Smagorinsky constant and the resulting eddy viscosity is nonnegative. Even when the Germano dynamic procedure is employed, a clipping of coefficients is normally required. However, it is well-acknowledged that the similarity model and the real SGS stress tensor generates large backscatter. Based on DNS databases for different turbulent flows \citep{Piomelli_1991,Domaradzki_1993,Domaradzki_1994,Domaradzki_1995}, when the cutoff $k_c$ is in the inertial subrange, the forward and inverse energy transfer characterized by the spectral eddy viscosity is comparable for $k<\frac{1}{2}k_c$. Beyond that, the forward energy cascade becomes more dominant when the LES cutoff is approached, justifying the above-mentioned strategy of providing higher energy dissipation at relatively large wavenumbers. To mimic the real actions of subgrid scales, ideally both the forward and the backward energy transfer should be included. Recall that the similarity model is simply the SGS stress defined on filtered scales, given that the `not-so-local' contributions between $k<ak_c$ and $k>k_c$ is neglected, so the inverse energy cascade has already been accounted for by the energy dissipation computed from equation (\ref{eq:epssfs}). Indeed, it has been established in a number of papers that the energy transfer beyond the energy-containing range is mostly local (see e.g. \citep{Domaradzki_1993,Domaradzki_1994,Domaradzki_1995,Domaradzki_2007b}). In practice, we found that the interscale model is sufficient to avoid numerical issues for all simulations in the current study, so an elimination of backscatter is not necessary and the desired distribution of energy dissipation can be retained.

To evaluate the direction of energy transfer in more detail, spatial distributions of instantaneous eddy viscosity for $Re_\tau=1000$ for the dynamic Smagorinsky model and the interscale model are shown in Fig. \ref{fig:Re1000nut}. In order to compare the two models more unambiguously, the eddy viscosities are computed based on a same snapshot of velocity field, which is taken from a relatively accurate simulation with the dynamic Smagorinsky model. Whether or not the viscous dissipation is partially removed has little impact on the contours thus is not shown. The distribution of eddy viscosity from the dynamic Smagorinsky model is more uniform: the majority of $\nu_t$ is within $[0, \, 5\times10^{-5}]$ and there are only very few regions with relatively high dissipation (mostly within the buffer layer). For our new model, the inverse energy transfer is apparently significant, leading to fairly large spatial variations.

\begin{figure}[t]
  \centering
    \includegraphics[width=0.6\textwidth]{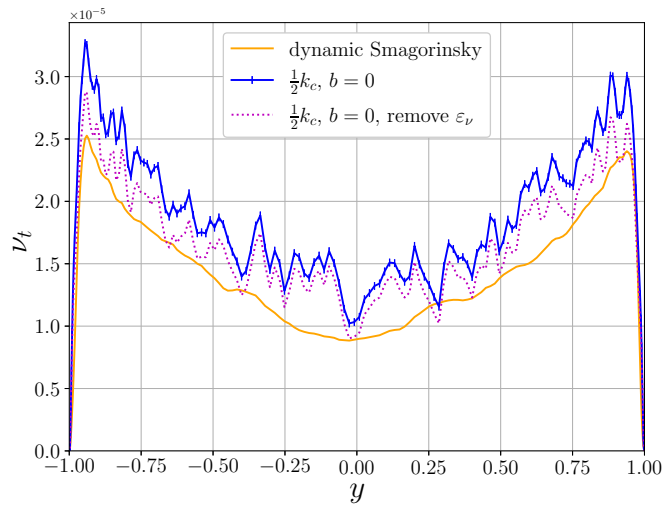}%
  \caption{Instantaneous eddy viscosity at $Re_\tau \approx 1000$ averaged in the horizontal directions.}
  \label{fig:Re1000nuts}
\end{figure}

Nevertheless, as one can observe from Fig. \ref{fig:Re1000nuts}, for the present model the eddy viscosities after averaging in homogeneous directions are positive, which explains why the method is numerically stable and robust in actual simulations. In addition, all models generate comparable eddy viscosities, consistent with the results for total SGS dissipation reported in Table \ref{tab:ct}. The inverse energy transfer for the interscale model originates from the similarity model, which correlates well with the real SGS stress in $a$ $priori$ analysis \citep{Liu_1994}. The allowance of backscatter is inherited from the similarity model, which is one of the key ingredients that differs the present model and other models that rarely take negative eddy viscosities into account. The recasting to the eddy viscosity form offers a strategy to provide an appropriate amount of total SGS dissipation while still maintaining the desired energy transfer distribution of the similarity model, so any further action is not necessary. Note that the procedure by itself also offers a capability of backscatter control, which will be further discussed in the next subsection.

\subsection{The effect of filtering and rescaling factor}
\label{sec:fil}

The spectral cutoff filter in the previous section offers a direct connection between the subgrid and subfilter scales, but it cannot be used for fully inhomogeneous flows. In this subsection, we explore the applicability of the new model based on a smooth filter. The Gaussian filter is chosen to compute the test-filtered field
\begin{equation}
  G(x, x') = \sqrt{\frac{\gamma}{\pi \Delta^2}} \exp\left(-\frac{\gamma|x - x'|^2}{\Delta^2} \right), \text{ or in spectral space } \widehat{G}(k) = \exp\left(\frac{-\Delta^2 k^2}{4\gamma} \right)
  \label{eq:f_gaus}
\end{equation}
where a common choice $\gamma=6$ is used for all simulations. The filter is smooth in both physical and spectral space. For the ease of implementation, in the current work the filter is only applied to the horizontal directions, in which $\Delta$ in the filtering kernel is replaced by the uniform mesh size in each direction. In practice, the Gaussian filter is more commonly used to compute the similarity model \citep{Bardina_1980,Piomelli_1988}, facilitating direct comparisons with our new approach. The similarity model computed from smooth filters also exhibits higher correlations in $a$ $priori$ analysis \citep{Liu_1994}. Moreover, as discussed in \citep{Vreman_1996,Thiry_2016}, a series expansion of the similarity model based on the Gaussian filter leads to
\begin{equation}
  \widetilde{\bar{u}_i \bar{u}_j} - \widetilde{\bar{u}}_i \widetilde{\bar{u}}_j = \frac{\Delta^2}{12} \frac{\partial \bar{u}_i}{\partial x_k} \frac{\partial \bar{u}_j}{\partial x_k} + O(\Delta^4),
  \label{eq:gm}
\end{equation}
in which the leading order term is the well-known gradient model (or the nonlinear part of the tensor-diffusivity model, sometimes also called the Clark model \citep{Clark_1979}). It is noteworthy that the factor $\frac{1}{12}$ can be changed if a different filter is chosen, so that the modeling may become suboptimal. Conversely, the connection with the gradient model justifies the current choice of the Gaussian filter. Akin to the similarity model, the gradient model is also known to have a good correlation with the SGS stress but leads to unstable simulation in practice \citep{Vreman_1996}. Because of that it is applied with zero-clipping \citep{Vreman_1997,Balarac_2013} or combined with the eddy viscosity model to generate mixed models \citep{Clark_1979,Fabre_2011}. Recasting the similarity model to the eddy viscosity form serves the same purpose.

\begin{figure}[t]
\subfloat
{%
  \centering
  \includegraphics[trim=60 0 210 0, clip, width=0.84\columnwidth]{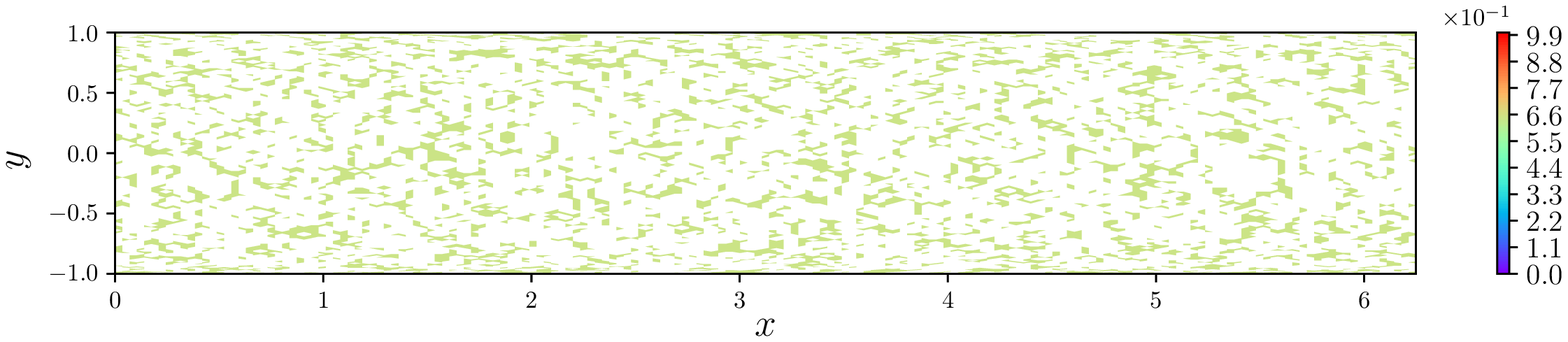}%
}
\put(-405,75){{\large $(a)$}}
\hfill
\subfloat
{%
  \centering
  \includegraphics[trim=60 0 210 0, clip, width=0.84\columnwidth]{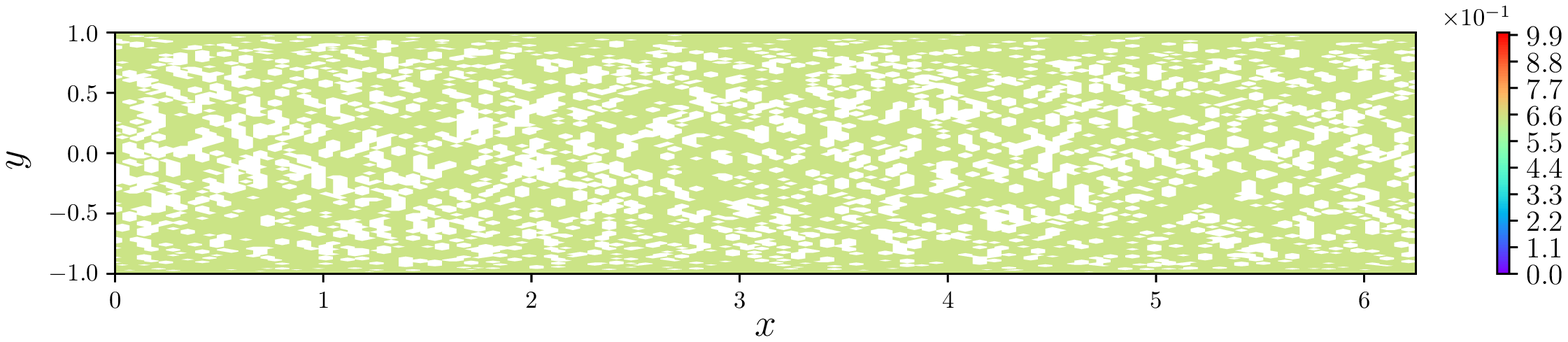}%
}
\put(-405,75){{\large $(b)$}}
\caption{Instantaneous SGS dissipation at $Re_\tau \approx 1000$, $(a)$ the similarity model, $(b)$ ITM. The colored region shows $\varepsilon_{\textsc{sgs}} \geq 0$, the white region indicates backscatter.}
\label{fig:Re1000eps}
\end{figure}

\begin{table}[ht]
  \centering
  \caption{\label{tab:gau} Results for ITM performed with the Gaussian filter. }
  \begin{tabular}{c|c|c|c|c}
  \hline
  \hline
$Re_\tau \approx 180$                   & Similarity & $b=0$   & $b=0.2$ & $b=0.4$ \\
\hline
$Err_{\textsc{dns}}$ (\%)               & 6.3125     & 2.4357   & 2.6815   & 3.7666   \\
$Err_{\text{m}}$ (\%)                   & 3.4405     & 1.8854   & 1.4155   & 1.1985   \\
$\varepsilon_{\textsc{sgs}}$           & 1.24E-04   & 1.01E-04 & 1.21E-04 & 1.44E-04 \\
$T_\nu$                                 & 1.09E-03   & 1.02E-03 & 1.01E-03 & 9.81E-04 \\
$\varepsilon_{\textsc{sgs}}/T_\nu$ (\%) & 11.3888   & 9.8714   & 12.0378 & 14.6404 \\
\hline
$Re_\tau \approx 1000$                 & Similarity & $b=0$   & $b=0.2$ & $b=0.4$ \\
\hline
$Err_{\textsc{dns}}$ (\%)               & 8.3044     & 3.2532   & 2.3428   & 2.4208   \\
$Err_{\text{m}}$ (\%)                   & 6.4558     & 3.8715   & 2.6519   & 1.2789   \\
$\varepsilon_{\textsc{sgs}}$           & 9.86E-05   & 1.25E-04 & 1.43E-04 & 1.67E-04 \\
$T_\nu$                                 & 6.27E-04   & 5.80E-04 & 5.53E-04 & 5.23E-04 \\
$\varepsilon_{\textsc{sgs}}/T_\nu$ (\%) & 15.7274   & 21.4656 & 25.8120 & 31.9689 \\
\hline
$Re_\tau \approx 2000$                 & Similarity & $b=0$   & $b=0.2$ & $b=0.4$ \\
\hline
$Err_{\textsc{dns}}$ (\%)               & 17.0285   & 9.5445   & 5.6619   & 4.3742   \\
$Err_{\text{m}}$ (\%)                   & 4.8101     & 2.2635   & 1.4963   & 0.4538   \\
$\varepsilon_{\textsc{sgs}}$           & 9.70E-05   & 1.33E-04 & 1.55E-04 & 1.86E-04 \\
$T_\nu$                                 & 4.77E-04   & 4.69E-04 & 4.40E-04 & 4.01E-04 \\
$\varepsilon_{\textsc{sgs}}/T_\nu$ (\%) & 20.3470   & 28.4099 & 35.2734 & 46.3981 \\
\hline
  \hline
  \end{tabular}
\end{table}

\begin{figure}[t]
\subfloat
{%
  \includegraphics[width=0.49\columnwidth]{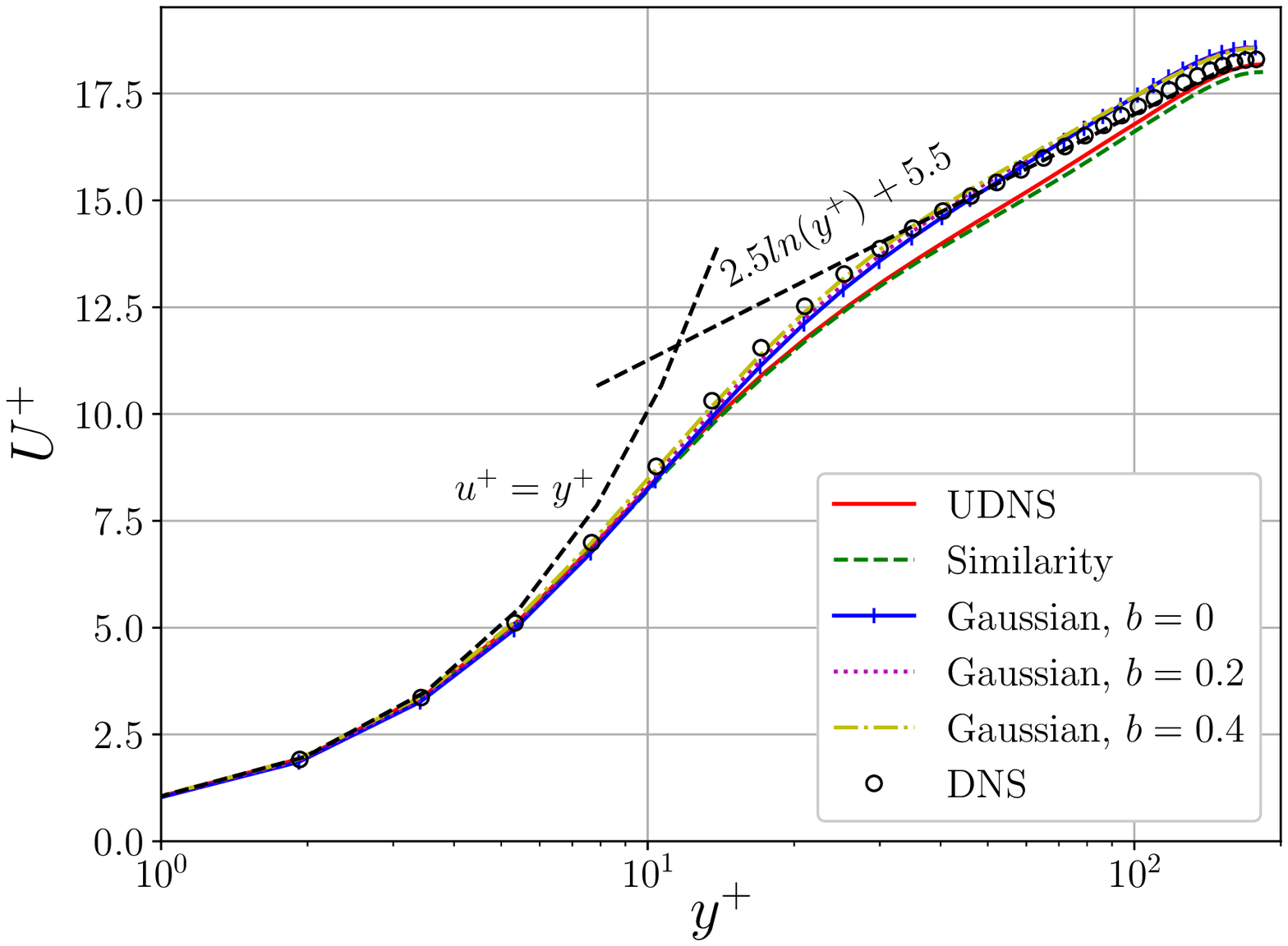}%
}
\subfloat
{%
  \includegraphics[width=0.49\columnwidth]{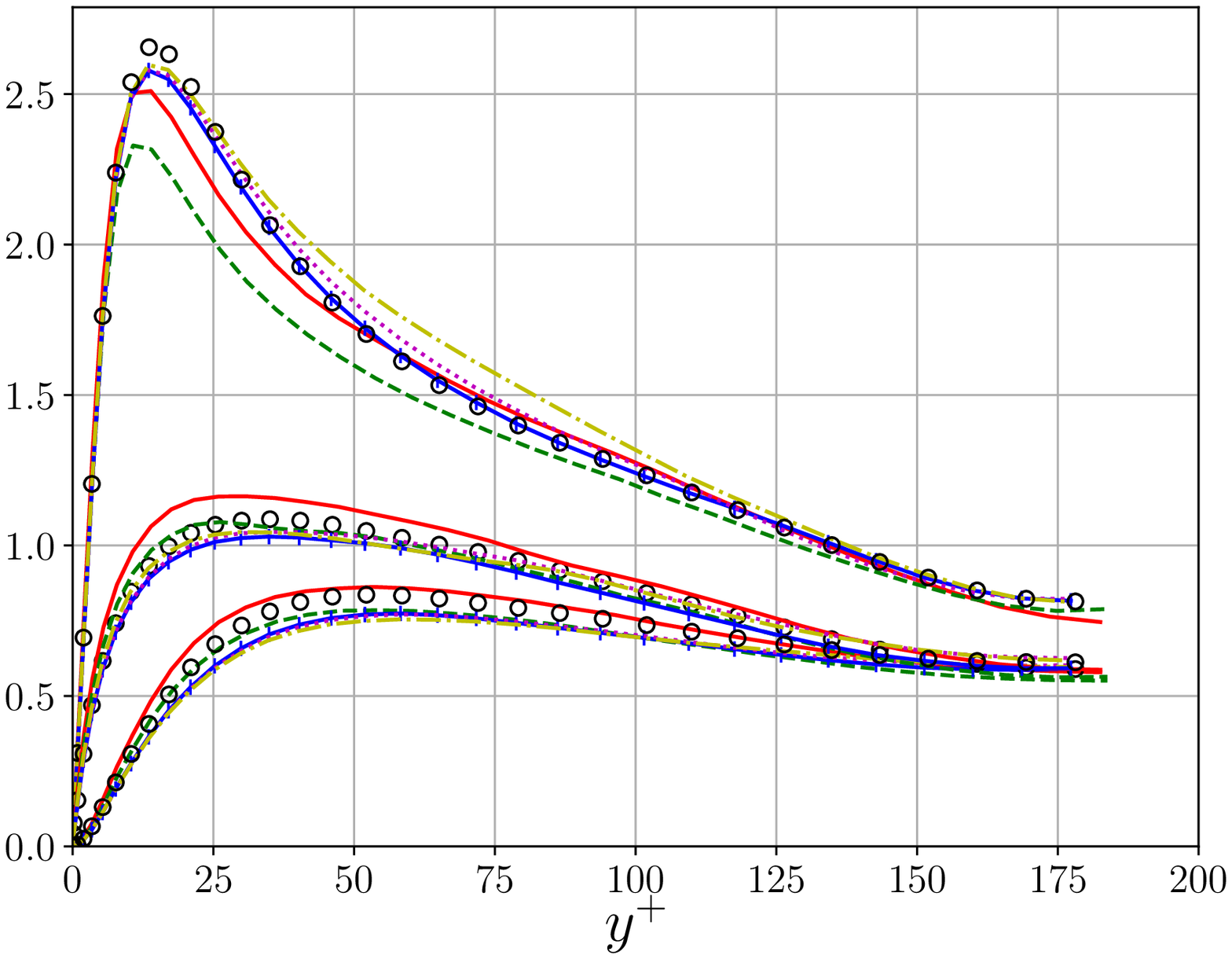}%
}
\caption{Mean velocity profiles (left) and rms velocities (right) for $Re_\tau=180$, the similarity model and ITM are performed with the Gaussian filter.}
\label{fig:Re180g}
\end{figure}

\begin{figure}[t]
\subfloat
{%
  \includegraphics[width=0.49\columnwidth]{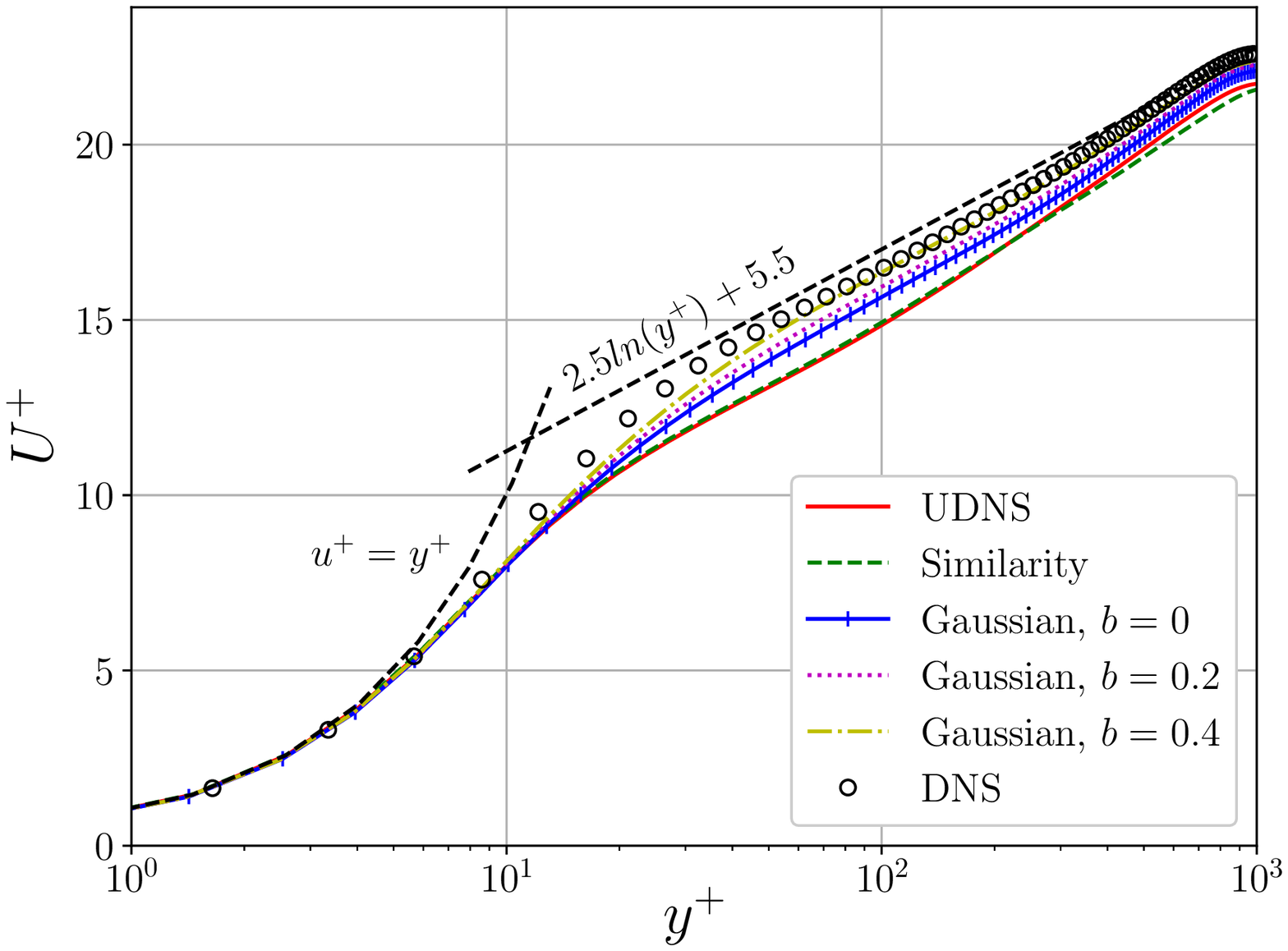}%
}
\subfloat
{%
  \includegraphics[width=0.49\columnwidth]{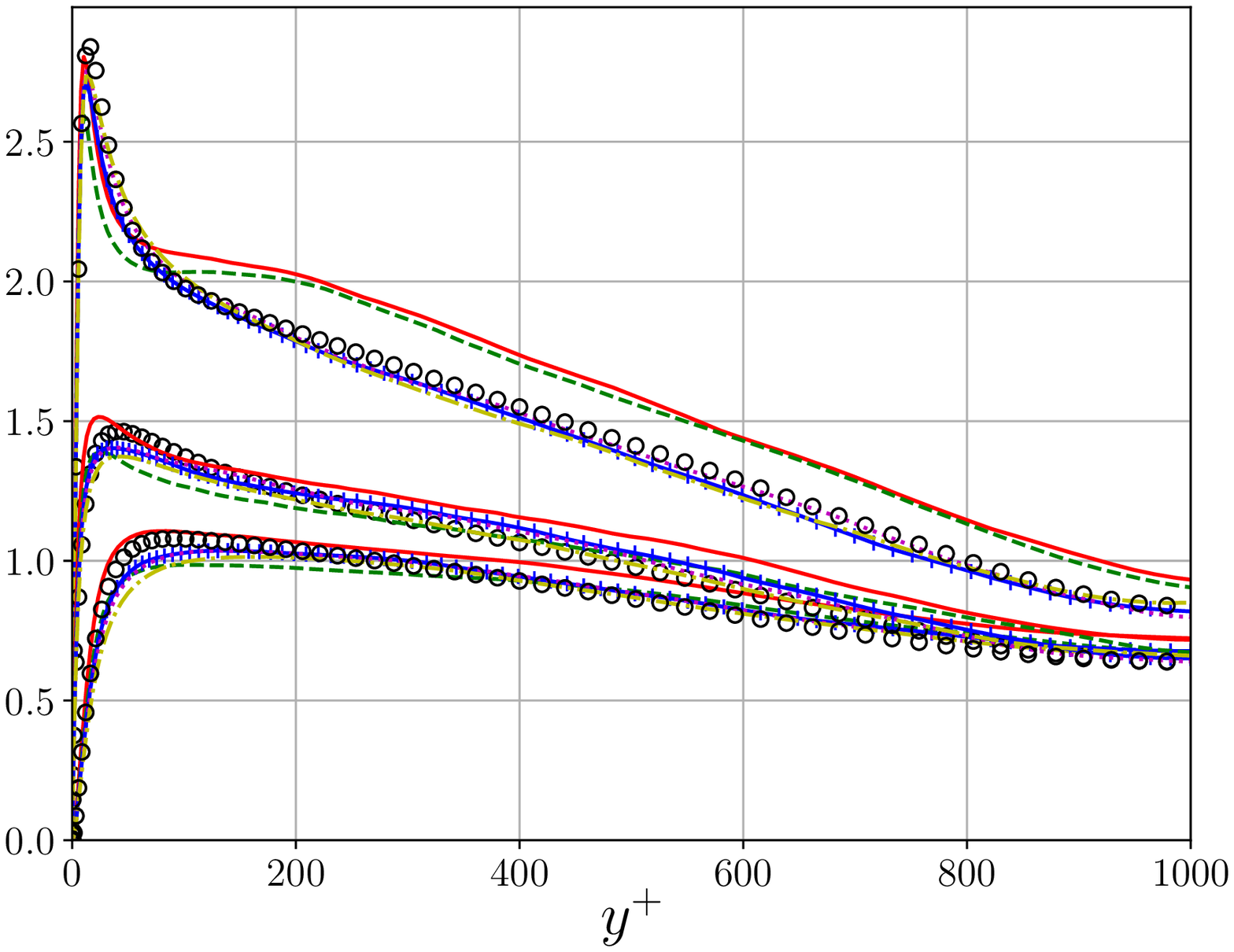}%
}
\caption{Mean velocity profiles (left) and rms velocities (right) for $Re_\tau=1000$, the similarity model and ITM are performed with the Gaussian filter.}
\label{fig:Re1000g}
\end{figure}

\begin{figure}[t]
\subfloat
{%
  \includegraphics[width=0.49\columnwidth]{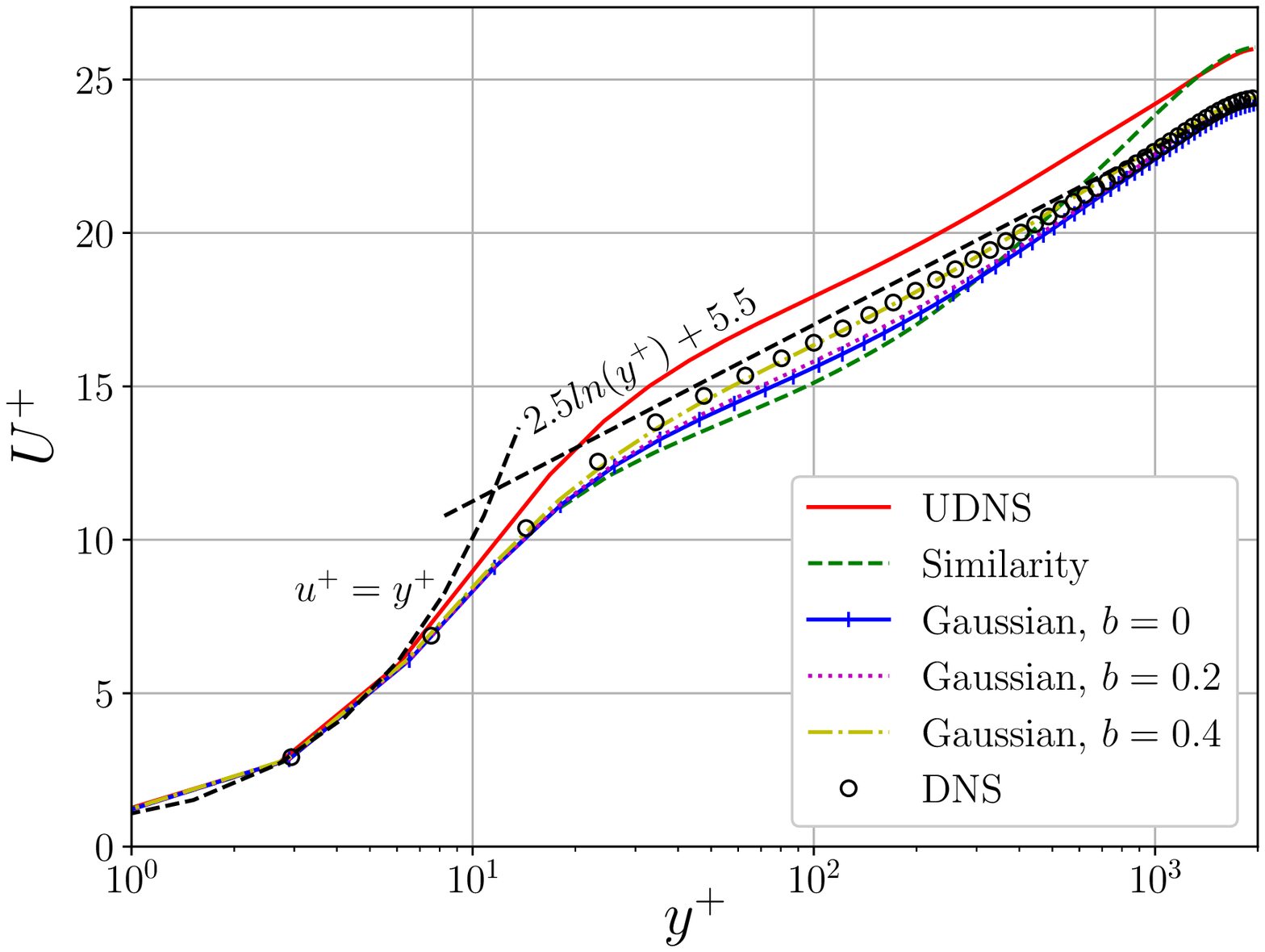}%
}
\subfloat
{%
  \includegraphics[width=0.49\columnwidth]{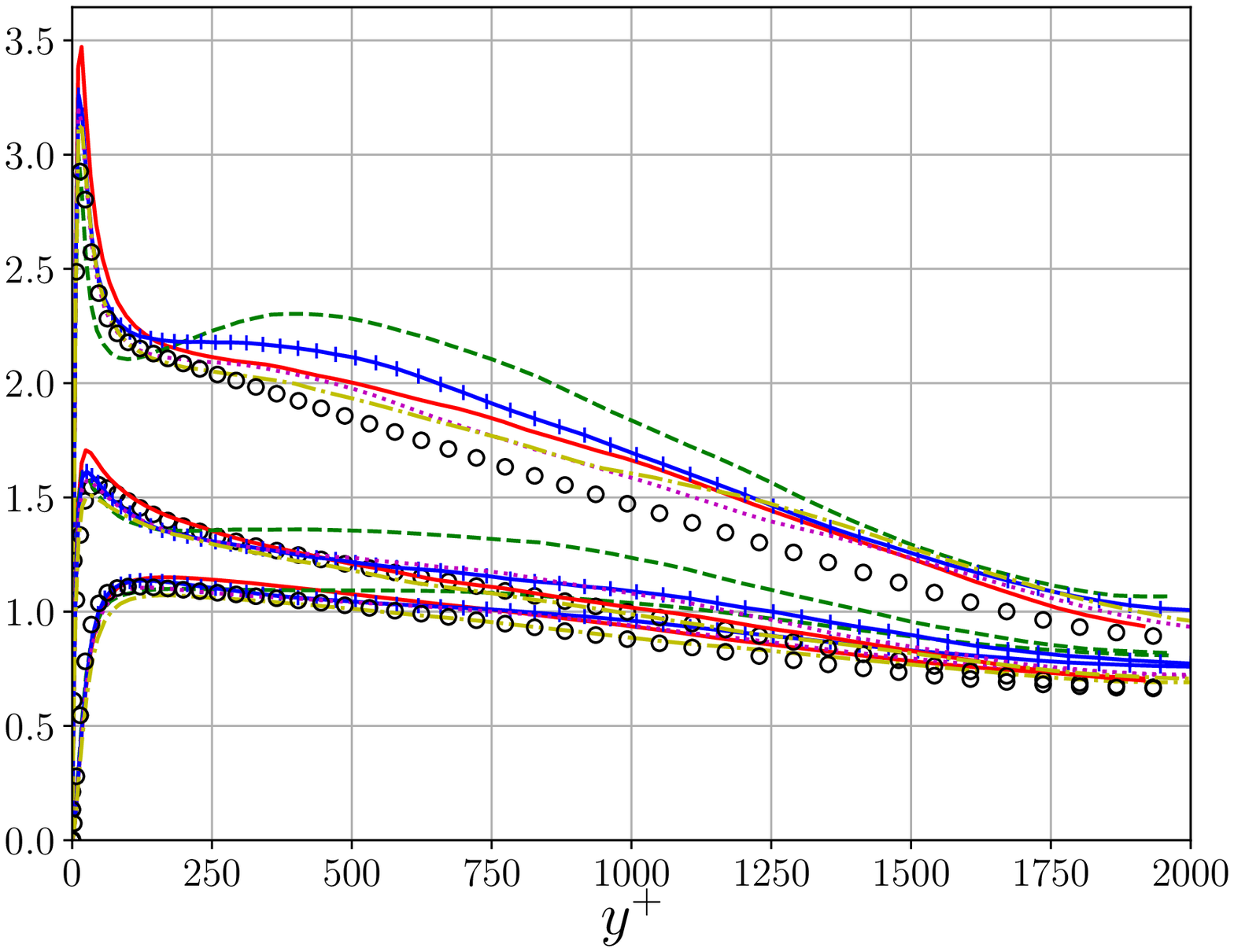}%
}
\caption{Mean velocity profiles (left) and rms velocities (right) for $Re_\tau=2000$, the similarity model and ITM are performed with the Gaussian filter.}
\label{fig:Re2000g}
\end{figure}

Following the discussion in the previous subsection, the effect of recasting to reduce the inverse energy transfer is evaluated in Fig. \ref{fig:Re1000eps}. Due to the relative wide range of quantities, we only show positive and negative energy dissipation in the plots. In the original similarity model \citep{Bardina_1980}, only roughly half of the SGS dissipation is positive, indicating the backscatter is significant. The observation is consistent with the previous studies for the SGS energy transfer \citep{Piomelli_1991,Domaradzki_1993}. After rescaling to the eddy viscosity form, the forward energy cascade occupies much more space. As outlined in the previous subsection, even though large backscatter is expected in fully resolved DNS runs, in LES with relatively low resolutions, the total SGS dissipation is positive and makes notable contribution to the combined effect of small scale energy removal. On the contrary, a large portion of energy production from the similarity model can easily destabilize simulations and generate unsatisfactory results. Therefore, the recasting in the current model can be considered as a simple but effective technique of backscatter control.

\begin{figure}[t]
\subfloat
{%
  \includegraphics[width=0.49\columnwidth]{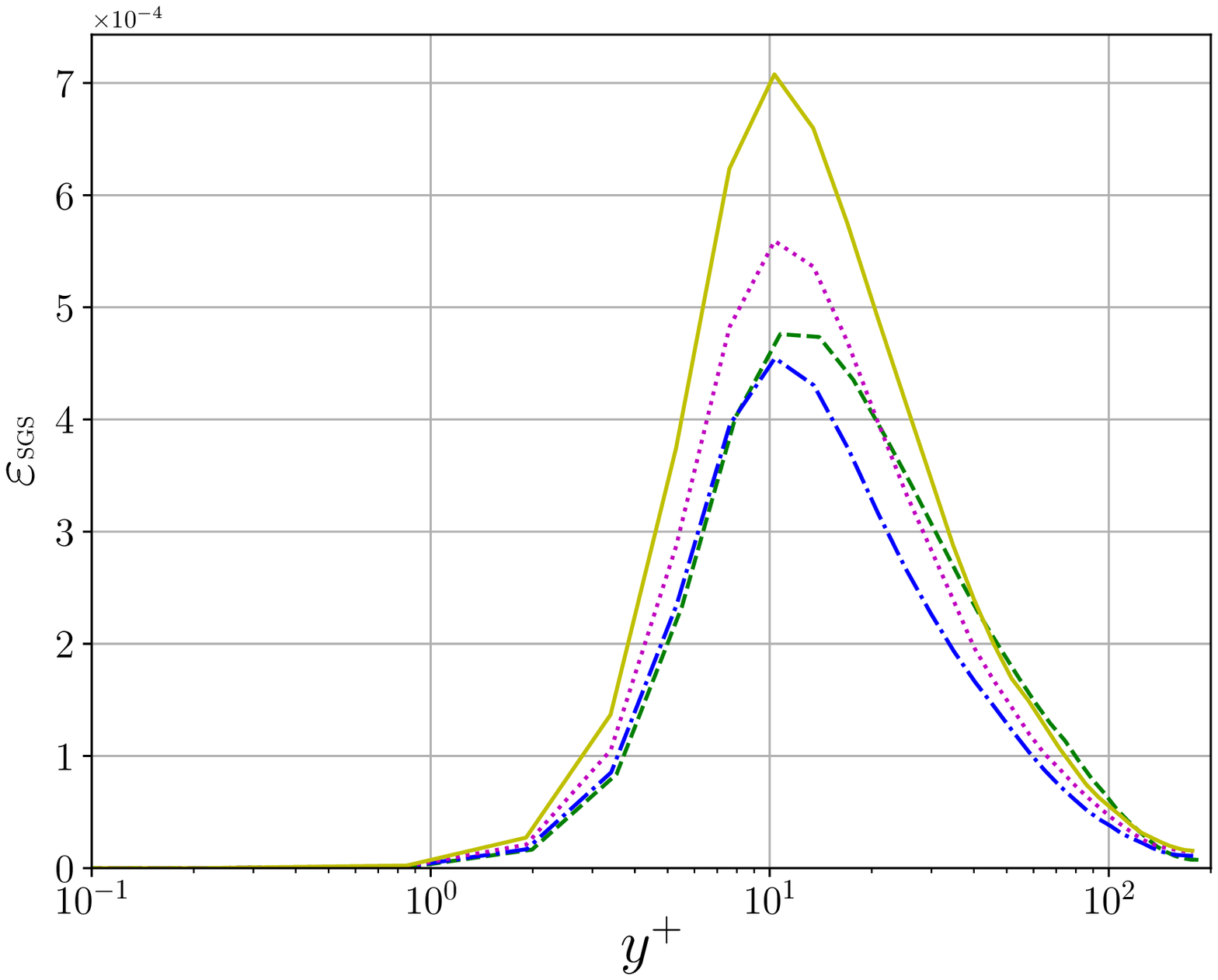}%
}
\subfloat
{%
  \includegraphics[width=0.49\columnwidth]{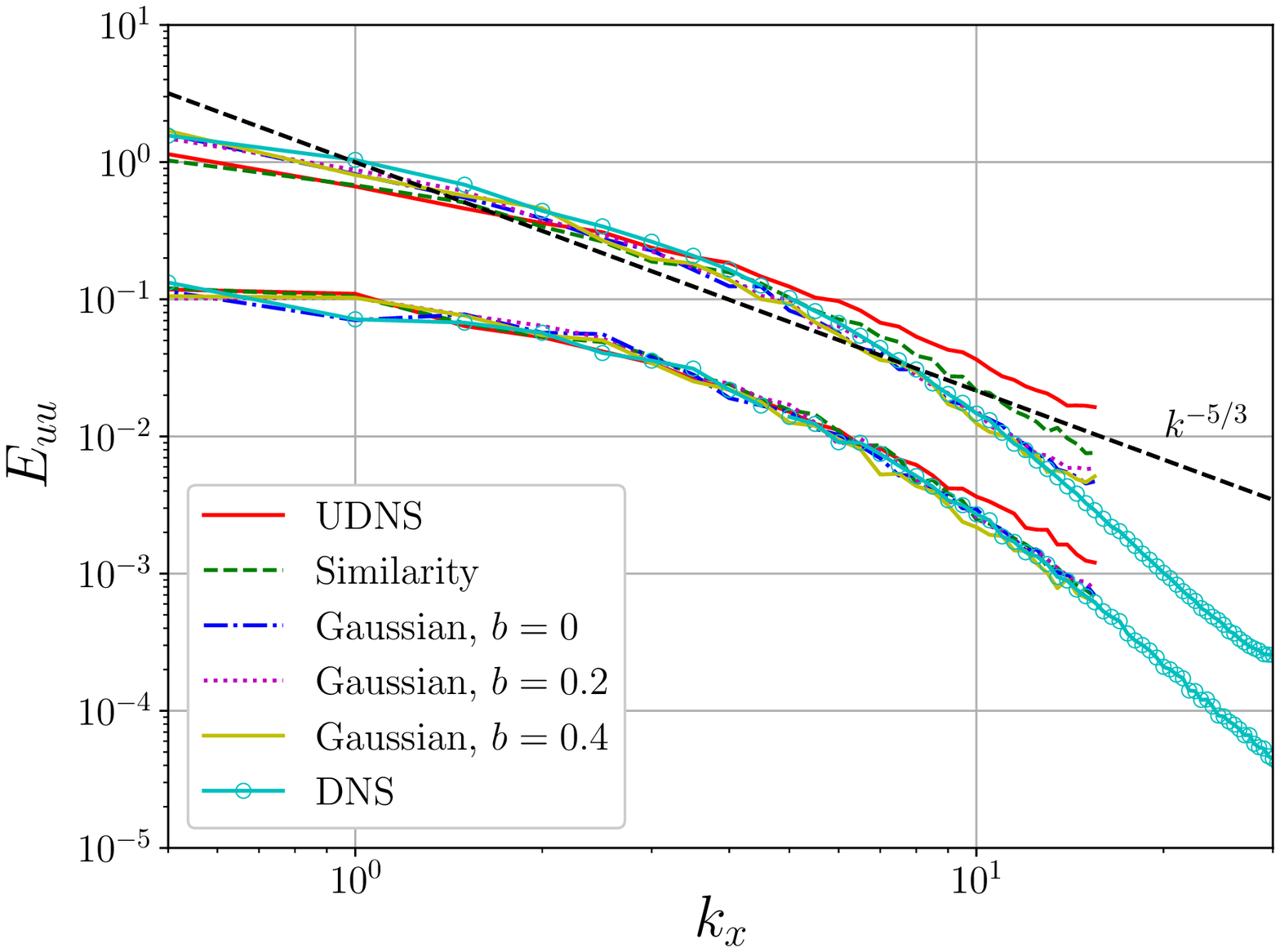}%
}
\caption{Horizontally averaged SGS dissipation (left) and 1-D streamwise energy spectra (right) for $Re_\tau=180$. The spectra in the upper part of the figure are for locations inside the buffer layer ($y^+ \approx 20$), and the spectra in the lower part are for locations close to the center ($y^+ \approx 180$). The ITM is performed with the Gaussian filter.}
\label{fig:Re180g1}
\end{figure}

\begin{figure}[t]
\subfloat
{%
  \includegraphics[width=0.49\columnwidth]{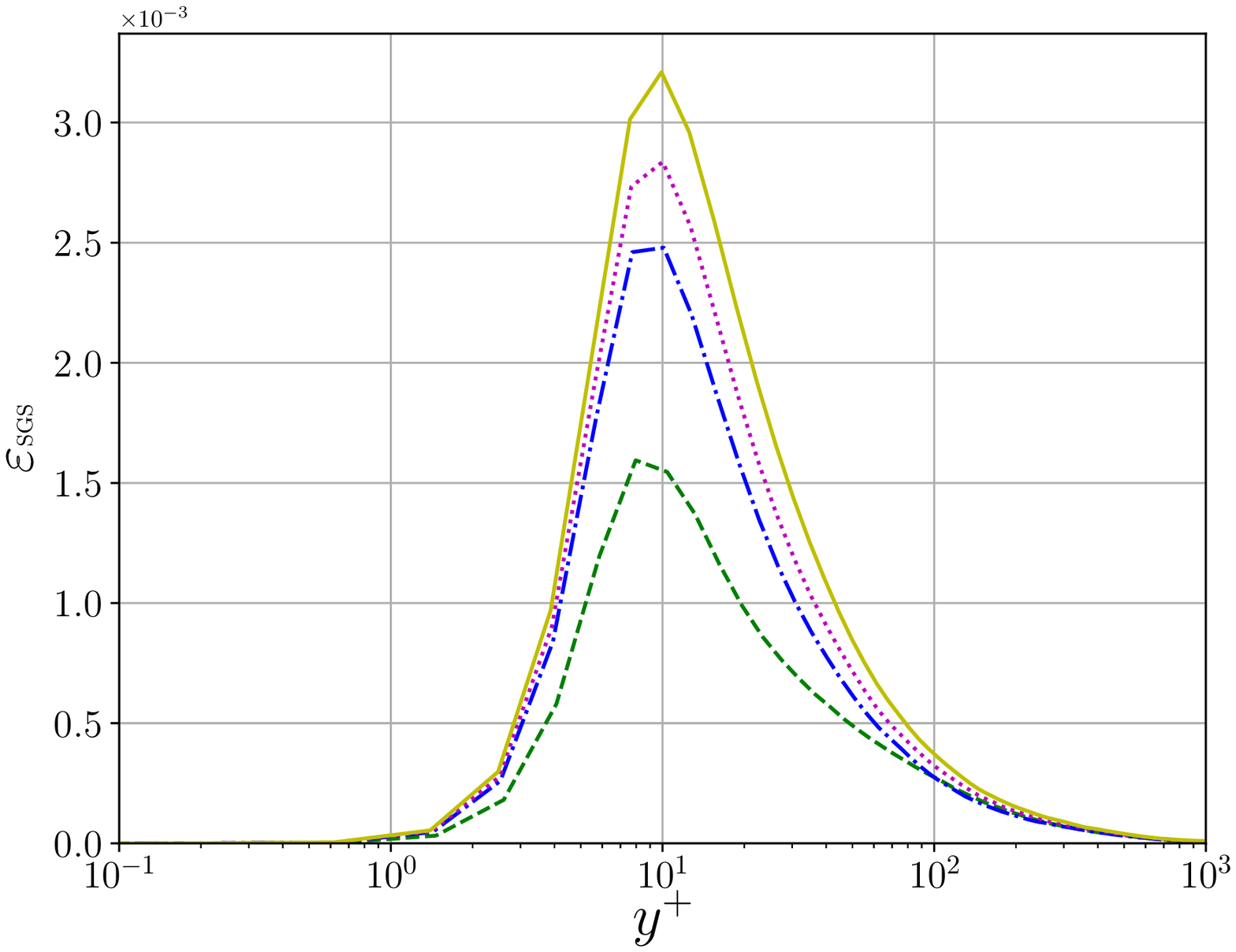}%
}
\subfloat
{%
  \includegraphics[width=0.49\columnwidth]{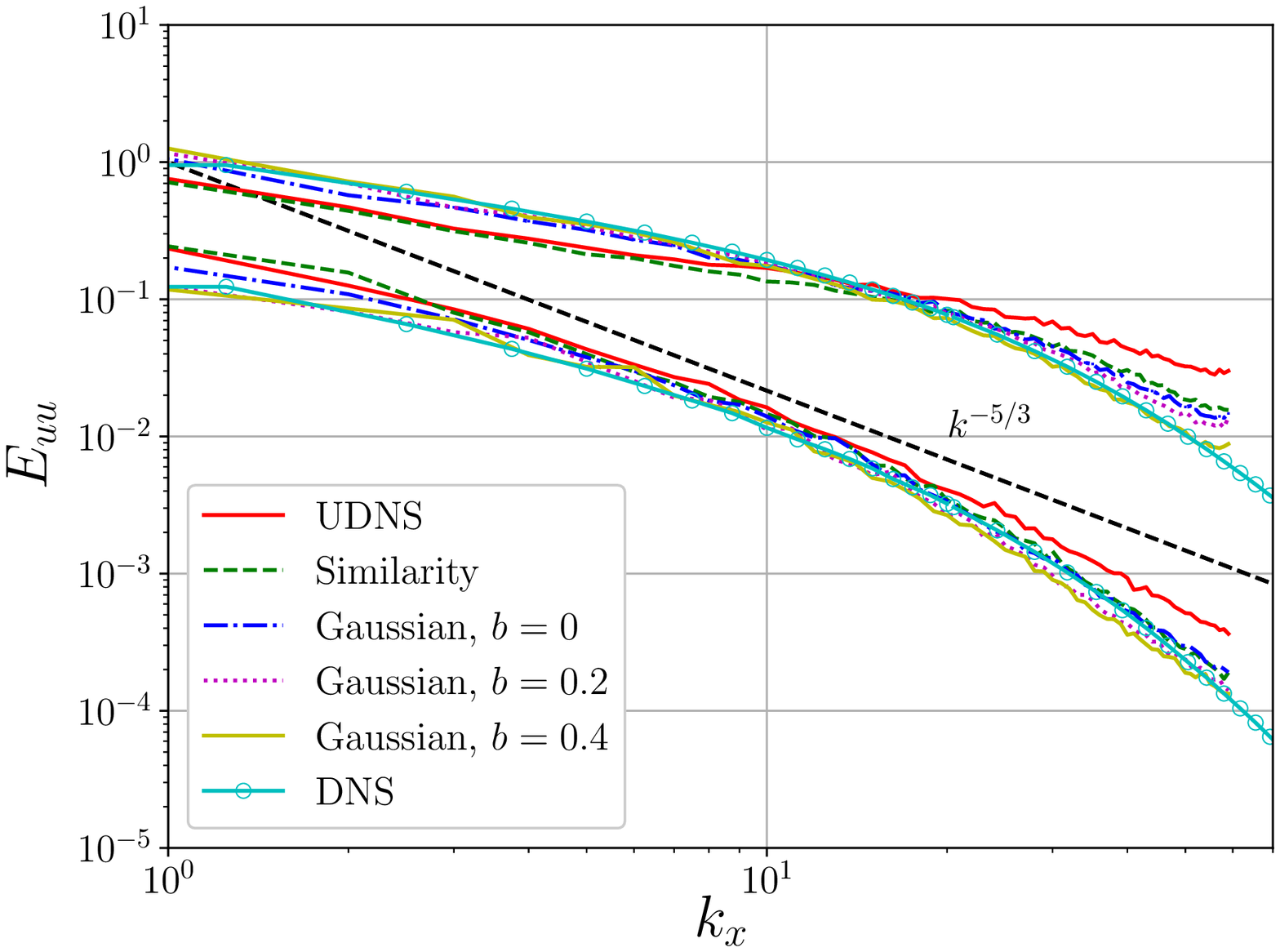}%
}
\caption{Horizontally averaged SGS dissipation (left) and 1-D streamwise energy spectra (right) for $Re_\tau=1000$. The spectra in the upper part of the figure are for locations inside the buffer layer ($y^+ \approx 20$), and the spectra in the lower part are for locations close to the center ($y^+ \approx 1000$). The ITM is performed with the Gaussian filter.}
\label{fig:Re1000g1}
\end{figure}

\begin{figure}[t]
\subfloat
{%
  \includegraphics[width=0.49\columnwidth]{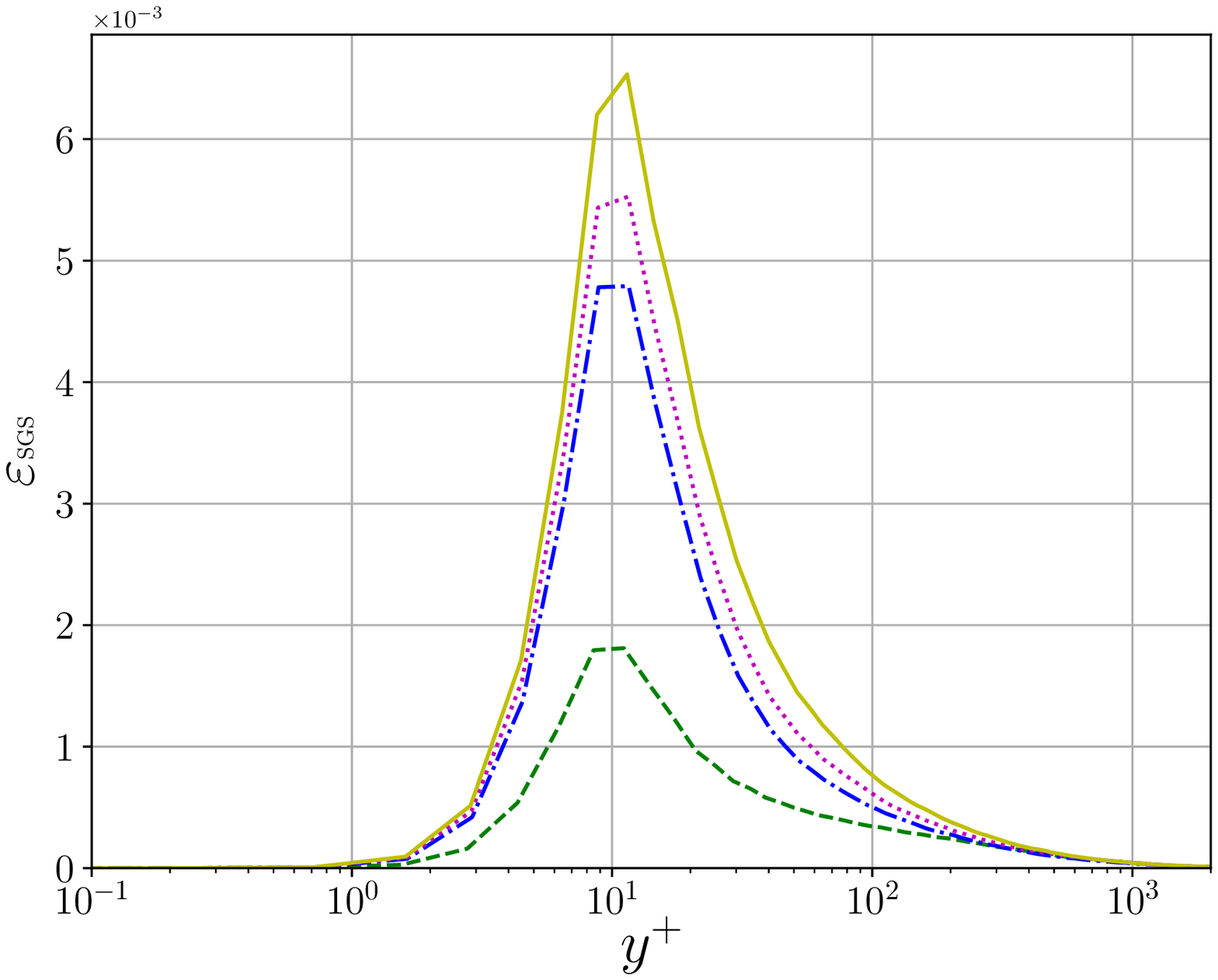}%
}
\subfloat
{%
  \includegraphics[width=0.49\columnwidth]{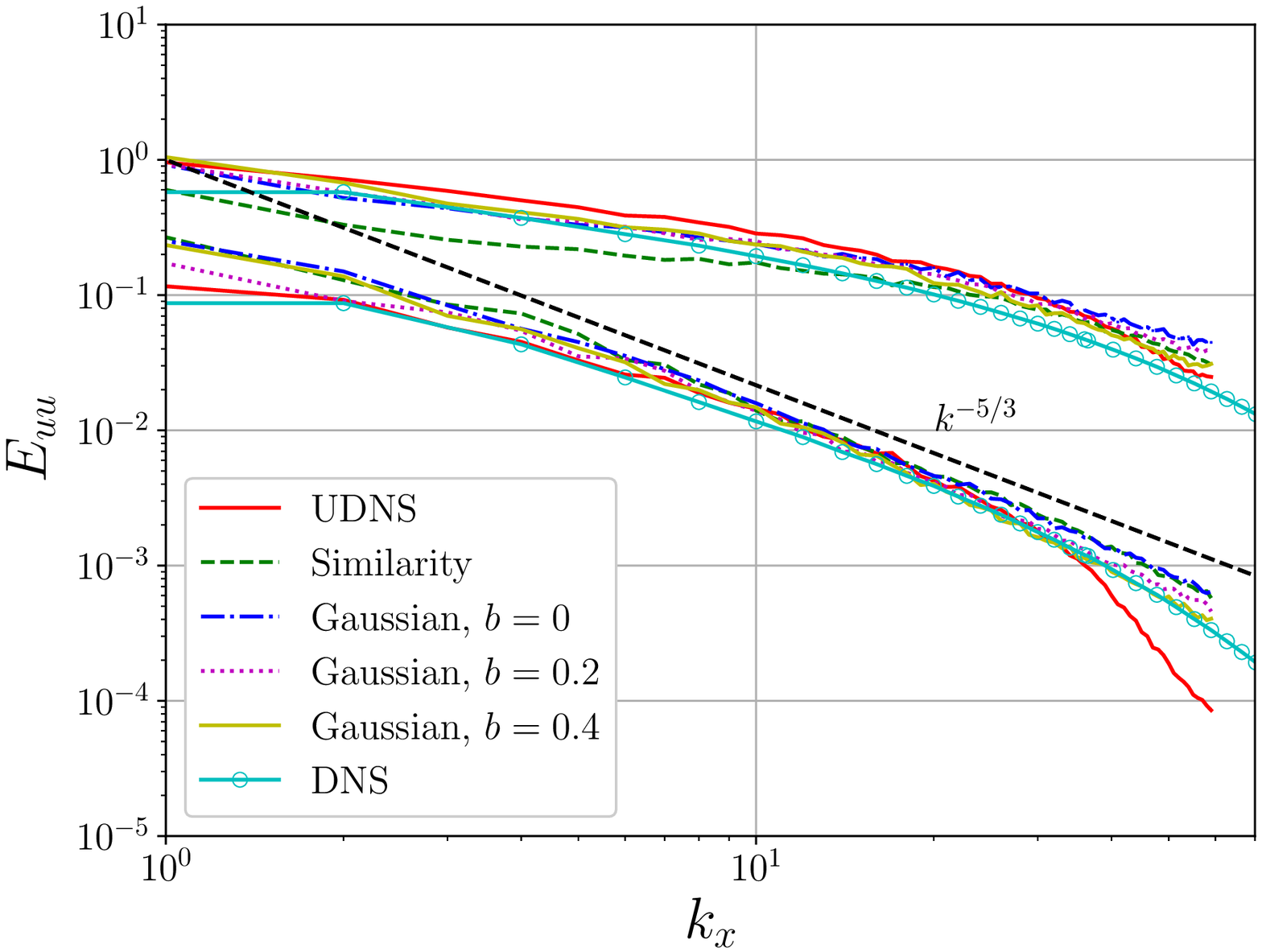}%
}
\caption{Horizontally averaged SGS dissipation (left) and 1-D streamwise energy spectra (right) for $Re_\tau=2000$. The spectra in the upper part of the figure are for locations inside the buffer layer ($y^+ \approx 20$), and the spectra in the lower part are for locations close to the center ($y^+ \approx 2000$). The ITM is performed with the Gaussian filter.}
\label{fig:Re2000g1}
\end{figure}

The recasting step with various factors $b$ in the present model is assessed in Table \ref{tab:gau} and Figs \ref{fig:Re180g}$\sim$\ref{fig:Re2000g}. As mentioned before, in previous works the similarity model is normally multiplied by a coefficient of $C_{\text{sim}} \approx 1$ \citep{Cook_1997,Meneveau_2000}, which is equivalent to $b \approx 0$ with $C_{\text{sim}} \approx 1$ in our new model. Compared with the ``UDNS'' case in Table \ref{tab:ct} the original similarity model does not offer any improvements. For ITM with $b=0$, the simulations become significantly more accurate. 
However, since the characteristic filtering width for the Gaussian filter (\ref{eq:f_gaus}) is larger than the sharp spectral filtering in the previous subsection, the total SGS dissipation with $b=0$ is likely inadequate. When the rescaling factor increases to 0.2 or 0.4, the low order statistics are reproduced more faithfully, especially for relatively high $Re$'s. One can also notice that the amount of total SGS dissipations in runs with $b=0.4$ are in general comparable to corresponding cases in Table \ref{tab:ct} with a different filter and the overall accuracy does not differ much. This indicates that the exact choice of filtering is also not crucial as long as the total SGS dissipation is estimated accurately.

The differences caused by different values of the rescaling factor $b$ are further evaluated based on the plane averaged SGS dissipation and the energy spectra in Figs \ref{fig:Re180g1}$\sim$\ref{fig:Re2000g1}. As expected, when $b$ increases, the SGS dissipation increases monotonically while maintaining roughly the same shape. The vertical distribution of $\varepsilon_{\textsc{sgs}}$ for the similarity model also looks similar, but the peak value becomes much lower for the two higher $Re$'s. This is likely due to the large inverse energy transfer discussed in the beginning of this subsection. As mentioned in Sec. \ref{sec:nm}, the mismatch between the smallest scales in LES and the test-filtered field makes the similarity model incapable of generating large energy dissipation near the LES cutoff $k_c$. Consequently, although $\varepsilon_{\textsc{sgs}}$ from the similarity model and the ITM with $b=0$ may not differ much in the very beginning, the former failed to generate a sufficient amount of small scale energy dissipation, leading to underprediction of the total SGS dissipation. Additionally, the efficiency of energy removal for different rescaling factors $b \in (0, 0.2 ,0.4)$ can be assessed from the energy spectra (the right subfigures in \ref{fig:Re180g1}$\sim$\ref{fig:Re2000g1}). A higher value of $b$ leads to overall lower energy spectra, especially close to the LES cutoff. For $b=0$ and 0.2, there still exists some notable energy accumulations near $k_c$, suggesting the rescaling factor should be further increased. The spectra for ITM with $b=0.4$ follows the DNS benchmark more closely, so its overall accuracy is also the best. Therefore, the shape of the energy spectra can be considered as an indicator to find an optimal choice of $b$. Note that the spectra for channel flow is steeper than the $-5/3$ scaling, offering additional difficulties in the theoretical prediction of an optimal rescaling factor. Nevertheless, it is likely possible to determine $b$ on the fly based on the information in a LES run, in a spirit of our previous works \citep{Tantikul_2010,Sun_2018} for adaptive solution filtering and the recent formalization of autonomous LES \citep{Domaradzki_2021b}. A dynamic evaluation of $b$ that varies in space and time for more complex flows could be an interesting future work as well.

\section{Conclusions}
\label{sec:conc}

A subgrid scale modeling based on the interscale energy transfer among resolved scales in LESs \citep{Domaradzki_2021a} is extended by incorporating implementations in physical space. Our primary motivation is to formulate a general modeling procedure by making use of the well-established theoretical foundations of the spectral energy transfer. The approach consists of two main steps, which in combination takes advantage of both the structural and the functional type of SGS modeling. Following the original idea of reconstructing the total SGS dissipation from a test-filtered field, the subfilter scale stress in physical space is used to obtain the energy dissipation defined on scales $k \leq ak_c$ with $a < 1$. The SGS dissipation is then cast in the form of eddy viscosity and rescaled to the LES cutoff $k_c$ via a global rescaling factor $b$, with a special focus on maintaining the ``true'' SGS dissipation up to the smallest Kolmogorov scale. Consequently, the model accomplishes a predominant objective in SGS modeling, which is widely acknowledged to be a necessary condition for accurate LES \citep{Meneveau_2000}.

The new model is easy to implement and offers several favorable features. The SGS stress for a test-filtered field is equivalent to the similarity model \citep{Bardina_1983}, whose high correlation with the real SGS term helps to maintain a desired distribution of energy transfer. The scale mismatch for the similarity model is fixed in the recasting step, which automatically leads to an effective backscatter control. Unlike typical eddy viscosity models that are forced to preserve the property of being purely dissipative, the inverse energy cascade in the present model does not jeopardize numerical stability and thus is retained. Moreover, the concept of recasting the physical SGS energy transfer into the eddy viscosity form \citep{Kraichnan_1976} is general, and works well for different filter choices. The theoretical analyses and satisfactory results for homogeneous, isotropic turbulence \citep{Domaradzki_2021a,Domaradzki_2021b} suggests that the present model should work at very high Reynolds numbers as well. In $a$ $posteriori$ analysis of LESs at relatively low resolutions, consistent with the previous studies for isotropic turbulence, the modeling procedure is capable of offering an appropriate amount of total SGS dissipation for a wall-bounded flow. As a result, the nonphysical energy accumulation at small scales is removed, leading to significantly improved low order turbulent statistics compared with UDNS or LES with the similarity model.

The current study offers a first step to explore the performance of this general modeling procedure for inhomogeneous turbulent flows. Simulations of turbulent channel flow show that the method may not be very sensitive to specific choice of filtering or the rescaling factor $b$, facilitating applications to flows in more complicated geometries. Note that to improve the numerical stability of LES with the current model, in this work a 1-D averaging is performed when computing the eddy viscosity. For fully inhomogeneous flow, proven techniques in the dynamic procedure such as local spatial averaging \citep{Cadieux_2015} or the Lagrangian averaging \citep{Meneveau_1996} could be feasible choices. Besides, an eddy viscosity that further enhances small scale dissipations (i.e. a cusp near $k_c$ in spectral space) is expected to make the model numerically more robust. On the other hand, the modeling concept does not rely on specific filters, those that are easier to implement for nonuniform grids can be used in place of the Gaussian filter. In summary, the present work made the interscale energy transfer model more general, which can be directly applicable to more complex flows after minor changes.




\bibliography{ref}

\end{document}